\documentclass[aps,eqsecnum,nofootinbib,superscriptaddress,showpacs]{revtex4}
\usepackage{amsmath}
\usepackage{amsfonts}
\usepackage{amssymb}
\usepackage[dvips]{graphicx}  % for graphics
\usepackage{amsmath,amsfonts}
\usepackage{color}  % for color
\usepackage[all]{xy}
\def\be{\begin{eqnarray}}
\def\ee{\end{eqnarray}}
\def\beq{\begin{equation}}
\def\eeq{\end{equation}}

\def\({\left (}
\def\){\right )}

\newtheorem{theorem}{Theorem}[section]
\newtheorem{lemma}[theorem]{Lemma}
\newtheorem{proposition}[theorem]{Proposition}
\newtheorem{corollary}[theorem]{Corollary}
\newtheorem{definition}{Definition}[section]
\newtheorem{remark}[theorem]{Remark}
\newtheorem{example}[theorem]{Example}
\newtheorem{conjecture}[theorem]{Conjecture}

\newcommand{\qed}{\nobreak \ifvmode \relax \else
      \ifdim\lastskip<1.5em \hskip-\lastskip
      \hskip1.5em plus0em minus0.5em \fi \nobreak
      \vrule height0.75em width0.5em depth0.25em\fi}
\newcommand{\qcd}{\begin{flushright} $\Box$ \end{flushright}}
\begin{document}
%%%%%%%%%%%%%%%%%%%%%%%%%%%%%%%%%%%%%%%%%%%%%%%%%%%%%%%%%%%%%
%%%%%%%%%%%%%%%%%%%%%%   Title Page   %%%%%%%%%%%%%%%%%%%%%%%
%\rightline{Preprint-MTM}
%\rightline{hep-th/0605224}

\title{
On the splitting problem for Lorentzian manifolds with an $\mathbb{R}$-action with causal orbits
}

\author{I.P. Costa e Silva\footnote{Visiting Scholar. Permanent address: Department of Mathematics,
Universidade Federal de Santa Catarina, 88.040-900 Florian\'{o}polis-SC, Brazil.}}
\email{pontual.ivan@ufsc.br}
\affiliation{Department of Mathematics,\\
University of Miami, Coral Gables, FL 33124, USA.}
\author{J.L. Flores}
\email{floresj@uma.es}
\affiliation{Departamento de \'Algebra, Geometr\'{i}a y Topolog\'{i}a,\\ Facultad de Ciencias, Universidad de M\'alaga \\ Campus Teatinos, 29071 M\'alaga, Spain}

\date{\today}

\begin{abstract}

We study the interplay between the global causal and geometric structures of a spacetime $(M,g)$ and the features of a given smooth $\mathbb{R}$-action $\rho$ on $M$ whose orbits are all causal curves, building on classic results about Lie group actions on manifolds described by Palais \cite{Palais}. Although the dynamics of such an action can be very hard to describe in general, simple restrictions on the causal structure of $(M,g)$ can simplify this dynamics dramatically. In the first part of this paper, we prove that $\rho$ is free and proper (so that $M$ splits topologically) provided that $(M,g)$ is strongly causal and $\rho$ does not have what we call {\em weakly ancestral pairs}, a notion which admits a natural interpretation in terms of ``cosmic censorship''. Accordingly, such condition holds automatically if $(M,g)$ is globally hyperbolic. We also prove that $M$ splits topologically if $(M,g)$ is strongly causal and $\rho$ is the flow of a complete {\em conformal Killing causal} vector field. In the second part, we investigate the class of {\em Brinkmann spacetimes}, which can be regarded as null analogues of stationary spacetimes in which $\rho$ is the flow of a complete {\em parallel null} vector field. Inspired by the geometric characterization of stationary spacetimes in terms of standard stationary ones \cite{JS}, we obtain an analogous geometric characterization of when a Brinkmann spacetime is isometric to a standard Brinkmann spacetime. This result naturally leads us to discuss a conjectural null analogue for Ricci-flat $4$-dimensional Brinkmann spacetimes of a celebrated rigidity theorem by Anderson \cite{anderson}, and highlight its relation with a long-standing 1962 conjecture by Ehlers and Kundt \cite{EK}. If true, our conjecture provides strong mathematical support to the idea that gravitational plane waves are the most natural ``boundary conditions'' at infinity for vacuum solutions of the Einstein equation modeling regions outside gravitationally radiating sources.

\end{abstract}
%
%\pacs{04.20.Dw;04.20.Gz;02.40.-k}
%
\maketitle

\section{Introduction} \label{0}

Let $(M^n,g)$ be a {\em spacetime}, i.e., a connected smooth time-oriented Lorentzian manifold ($n \geq 2$). It is well-known that the time-orientation automatically ensures the existence of a smooth $\mathbb{R}$-action $\rho$ on $M$ with causal orbits. In general, the action $\rho$ can be very hard to describe. In fact, the isometry group of even compact Lorentzian manifolds can be non-compact, which gives a very rich dynamical structure to the orbits \cite{zeghib,piccionezeghib}, actually richer than is the case for Riemannian manifolds. However, it turns out that certain simple restrictions on the {\em causal structure} of $(M,g)$ can place definite constraints on any such action, which in turn may come to bear on the global geometry of $(M,g)$. Such interplay may be more keenly felt if the action has previous interesting additional features, e.g., if it is isometric. It is interesting to point out that none of the causality restrictions we consider here apply to compact spacetimes, which are not even chronological \cite{BE,oneill}. There arises here, therefore, a curious situation in which non-compact manifolds can be much easier to analyze than compact ones.

As an example of direct interest to us here, consider the case when the $\mathbb{R}$-action $\rho$ is the flow of a smooth complete timelike and {\em Killing} [resp. {\em conformal Killing}] vector field $X:M \rightarrow TM$. In this case, $(M,g)$ is said to be {\em stationary} [resp. {\em conformastationary}] (with respect to $\rho$). In this context Harris \cite{Harris} showed that if $(M,g)$ is {\em chronological}, i.e., has no closed timelike curves, then the action generated by $X$ is proper and free. Therefore, $M$ has the structure of a trivial principal $\mathbb{R}$-bundle over the orbit space $M/\mathbb{R}$, that is, $M$ is diffeomorphic to $\mathbb{R} \times M/\mathbb{R}$, a situation which we will informally refer to by saying that $M$ {\em splits topologically} with respect to the action.

The topological splitting of stationary spacetimes appears extensively in applications. One of the reasons for its interest is that in this case there exists a Riemannian metric induced on the orbit space $M/\mathbb{R}$ whose geometric structure is directly related to that on $M$. This fact had already been exploited by Geroch \cite{geroch1,geroch2}, even prior to the more rigorous Harris' result, to induce a system of elliptic differential equations on $M/\mathbb{R}$ equivalent to the Einstein field equation on $(M,g)$ for a given energy-momentum tensor $T$. Geroch also showed how these equations could be used to generate new solutions to the Einstein equations. This so-called {\em projection formalism} on $(M,g)$ was instrumental, e.g., in the proof of the uniqueness theorems for electrovac spacetimes containing a black hole (see, e.g., \cite{heusler} and references therein).

Stationary spacetimes topological splitting of the sort described above can accompanied by an interesting concomitant decomposition or splitting of the metric $g$. It is well-known that any stationary spacetime is {\em locally} isometric to a {\em standard stationary spacetime}, that is, a product manifold $M = \mathbb{R} \times S$ with Lorentzian metric $g$ given by
\begin{equation}
\label{stationarymetric}
g_{(t,x)} = -\beta(x) dt^2 + 2 \omega_x dt + \overline{g}_x,
\end{equation}
where $\beta, \omega$ and $\overline{g}$ are respectively a smooth positive function, a smooth 1-form and a smooth Riemannian metric on $S$. The timelike vector field $X$ generating the action $\rho$ is locally identified with $\partial_t$. Javaloyes and S\'{a}nchez \cite{JS} have shown that this local splitting can be made global if and only if $(M,g)$ is {\em distinguishing}, a condition stronger than causality but weaker than strong causality in the so-called causal ladder \cite{MS} of spacetimes. (Actually, the results in \cite{JS} are more general and apply to {\em conformastationary} spacetimes, but this description suffices to our purposes here.) Standard stationary spacetimes are much more amenable to a detailed analysis than general stationary spacetimes, since they have, among other things, an associated Finsler structure \cite{caponio}.

This paper is devoted to investigating the splitting problem for a spacetime $(M,g)$ with a smooth action $\rho$ given by the flow of a complete {\em causal} vector field $X$, i.e., $g(X,X) \leq 0$ and $X$ everywhere nonzero, but not necessarily timelike or (conformal) Killing. As we have hinted above, we will {\em not} be interested here in exploring the dynamical richness of the action $\rho$ alluded to before, but will only be concerned, on the one hand, with giving some natural criteria on the causality of $(M,g)$ to ensure that $\rho$ is free and proper, so that $M$ splits topologically; and, on the other hand, with obtaining an associated metric splitting in the particular when $X$ is null (i.e., $g(X,X) =0$ and $X$ is everywhere non-zero).

Let us briefly describe here our main results. Our principal topological splitting results are Theorems \ref{strongcausality} and \ref{jose} in Section \ref{0.3}. Theorem \ref{strongcausality} states that the action is indeed free and proper provided that $(M,g)$ is strongly causal,
%\footnote{In spite of its name, this condition is a relatively mild constraint on the causal structure of $(M,g)$.},
and that this action does not have what we call a {\em weakly ancestral pair}. This condition, which is defined precisely in Section \ref{0.3}, is a direct adaptation of a notion introduced by Harris and Low \cite{HL}, and has a natural interpretation in physical contexts as a form of ``cosmic censorship'' (cf. the remark just before Proposition \ref{noancestral}). Indeed, this condition is automatically satisfied when $(M,g)$ is globally hyperbolic, and hence {\em every} complete causal vector field generates a proper free action for the latter class of spacetimes. (This particular result, which is implicit in previous treatments \cite{JS,Josetal}, is explicitly stated here as Corollary \ref{GH}.) Theorem \ref{jose} applies specifically to the case of a {\em (conformal) isometric} action with causal orbits: in this case, strong causality ensures by itself that $(M,g)$ splits topologically, without the assumption on weakly ancestral pairs. When specialized to the purely timelike context, our methods allow us to give somewhat different proofs of some of Harris' \cite{Harris} and Harris and Low's \cite{HL} results (see Theorems \ref{harris} and \ref{harris2} below).

We also obtain a global rigidity result for the special case when the vector field $X$ with flow $\rho$ is {\em null} and {\em parallel} (i.e., $\nabla X=0$; here and hereafter, $\nabla$ denotes the Levi-Civita connection of $g$). Lorentz manifolds admitting such a vector field are called {\em Brinkmann spaces} and are known to have special Lorentzian holonomy (see, e.g., \cite{batat,baum,galaev,leistner} and references therein for recent results), and as such have a number of interesting geometric properties. A classic result by Brinkmann \cite{brinkmann} establishes that these spacetimes have local coordinates $(u,v,x^1,\ldots, x^{n-2})$ (now known as {\em Walker coordinates} after \cite{walker}) on some open subset $U \subseteq M$ for which the line element of $g$ is given by
\begin{equation}
\label{localBrink}
ds^2|_{U} = 2du(dv + \Omega_i(u,x)dx^i) + H(u,x)du^2 + \gamma_{ij}(u,x)dx^idx^j
\end{equation}
where $\Omega := \Omega_idx^i$ is a one-form and $\gamma:= \gamma_{ij}dx^idx^j$ is a metric, both defined on $U$, and $X$-invariant, and $X|_{U} = \partial_v$. Note the similarity with stationary spacetimes and their local standard form. This naturally motivates the question: when is a spacetime $(M,g)$ possessing a null parallel vector field $X$ {\em globally} of the form (\ref{localBrink})? If this is the case, in analogy with the timelike case, we shall refer to $(M,g)$ as a {\em standard Brinkmann spacetime}, for which a more precise definition is given in Section \ref{0.5} below (cf. Definition \ref{BEspacetimes}). Our main result for this part implies that a causal Brinkmann spacetime is a standard Brinkmann spacetime if and only if the action has a slice, i.e., hypersurface which is intersected exactly once for each orbit, and satisfies a technical condition to be defined more precisely in Section \ref{0.5}. We then discuss in which sense this can be regarded as an analogue of the Javaloyes-S\'{a}nchez result for stationary spacetimes.

%Again, the importance of such global splitting is to facilitate concrete applications. Standard Brinkmann spacetimes have been much studied in the physical literature since the striking discovery by Eisenhart \cite{eisenhart} that the description of particles' trajectories in Newtonian Physics can be completely given in terms of geodesics in such spacetimes (see \cite{minguzzi1,minguzzi2} and references therein for further results in that vein). This fact is a consequence of the special holonomy of such spacetimes, which allows one, under favorable circumstances, to obtain a reduction of the frame bundle of $M$ to a principal subbundle having the Galilei group as structure group \cite{duval1, duval2}, which in turn is the symmetry group of Newtonian Physics.

On the other hand, if a standard Brinkmann spacetime $(M,g)$ is such that $\Omega \equiv 0$ and $h_{ij} \equiv \delta_{ij}$ globally, then $(M,g)$ is called a {\em generalized pp-wave} \cite{BE,EK}. This family of spacetimes contains in particular the important class of {\em gravitational plane waves}, and provide an idealized description of gravitational waves in General Relativity. In this ambient, our global splitting result for Brinkmann spaces is a first step for the proof of a rigidity result analogous to the remarkable theorem obtained by Anderson in \cite{anderson}. In fact, in that paper Anderson uses the global splitting result for stationary spacetimes to establish that {\em every geodesically complete, chronological, Ricci-flat $4$-dimensional stationary spacetime is isometric to (a quotient of) flat Minkowski spacetime}. Now, recall that gravitational plane waves are geodesically complete vacuum spacetimes which are non-flat, and even though non-stationary in general, they still admit a null Killing (indeed parallel) vector field. By analogy to Anderson's rigidity result, it becomes natural to ask if they are the only spacetimes satisfying these properties. We discuss this issue in more detail in Section \ref{0.6}

%More precisely, a conjecture of the following general form becomes natural: any {\em strongly causal, Ricci-flat, $4$-dimensional Brinkmann spacetime satisfying certain completeness hypothesis must be (a quotient of) a gravitational plane wave.} We

%(Clarifying just which completeness condition would work in this null context is part of the problem; it is unclear to us if geodesic completeness would suffice.) The global splitting result we prove in this paper effectively reduces the general problem to the particular case of a standard Brinkmann spacetime. However, it turns out that under the relevant hypotheses, a standard Ricci-flat Brinkmann space is actually a pp-wave \cite{CFH}. Thus, our proposed conjecture will be true if one establishes another, much older conjecture posed by Ehlers and Kundt in \cite[Section 2-5.7]{EK}, which states that {\em every geodesically complete, Ricci-flat $4$-dimensional pp-wave is a plane wave} (see, for instance, \cite{B}, \cite{HR}, \cite{FlS}, \cite{leistner}, \cite{CFH} for some discussion and partial solutions to this problem). In particular, we establish elsewhere \cite{CFH} a particular version of our extended conjecture.

The rest of this paper is organized as follows. In Section \ref{0.1}, we present some generalities on group actions, mainly to establish terminology and make the paper reasonably self-contained. In Section \ref{0.3} we discuss the interplay between causal conditions on $(M,g)$ and the action of a causal vector field thereon and obtain some topological splitting results. Specifically, our purpose therein is to establish a number of topological splitting results based on suitable causal assumptions on $(M,g)$. We present a brief review of the Javaloyes-S\'{a}nchez splitting for stationary spacetimes in Section \ref{0.4} in a form such as to serve as a template for our own rigidity results for the geometry of Brinkmann spacetimes, described in detail in Section \ref{0.5}. Finally, we formulate the conjecture indicated above about gravitational plane waves and discuss its main aspects in Section \ref{0.6}.
%%%%%%%%%%%%%%%%%%%%%%%%%%%%%%%%%%%%%%%%%%%%%%%%%%%%%%%%%%%%%
\section{Generalities on group actions: preliminaries \& notation}\label{0.1}
%%%%%%%%%%%%%%%%%%%%%%%%%%%%%%%%%%%%%%%%%%%%%%%%%%%%%%%%%%%%%

In this section, in order to establish the terminology and notation we shall use in later sections, we review those aspects of the basic theory of group actions which will have applicability to our approach of the problem of splitting as presented in this paper. We follow closely Palais' classic treatment on the subject given in Ref. \cite{Palais}, to which we refer the reader for the proofs. Accordingly, for the sake of clarity we indicate, for each definition, proposition, etc., their counterparts in that reference.

However, although many of the results therein have wider applicability, for the sake of brevity we shall not preserve the level of generality of \cite{Palais}. Specifically, throughout this paper, we fix a smooth $n$-dimensional manifold $M$, a Lie group $G$\footnote{Throughout this paper, let it be understood that {\em smooth} means $C^{\infty}$ and that manifolds, and in particular Lie groups, are always assumed to be Hausdorff, locally compact topological spaces with a countable basis, and therefore also paracompact.}, and a smooth action $\rho:G \times M \rightarrow M$. (Palais \cite{Palais} calls $(M,\rho)$ a {\em differentiable $G$-space}.) If there is no risk of confusion, we shall often drop an explicit mention of $\rho$, refer to $M$ as a $G$-space and write $gp :=\rho(g,p)=\rho_g(p)$, $\forall p \in M, \forall g \in G$.

For any set $U \subseteq M$, and any $g \in G$, we write $gU := \{gp \, : \, p \in U \}$, and $Gp:=\{gp \, : \, g \in G\}$ for the orbit of $p \in M$. $G_p$ denotes the isotropy group at $p$, i.e. $G_p =\{ g \in G \, : \, gp =p \}$. The action $\rho$ is {\it free} if $G_p = \{e\}$ for every $p \in M$. Following the notation in Ref. \cite{Palais}, for any sets $U,V \subseteq M$, we define
\[
((U,V)) := \{ g \in G \, : \, gU \cap V \neq \emptyset \}.
\]

Finally, we shall denote the quotient space of $M$ by the action (with the quotient topology) as $M/G$, and by $\pi: M \rightarrow M/G$ the (continuous) standard projection. Note that $\pi$ is an open map.

\begin{definition}[Def. 1.1.1 of \cite{Palais}]
\label{thin}
If $U$ and $V$ are subsets of $M$, then we say that $U$ is {\em thin relative to} $V$ if $((U,V))$ has compact closure in $G$. If $U$ is thin relative to itself, then we say that $U$ is {\em thin.}
\end{definition}

Since $gU \cap V =g(U \cap g^{-1}V)$, it follows that if $U$ is thin relative to $V$, then $V$ is thin relative to $U$, and so we often say that $U$ and $V$ are {\em relatively thin}. It is easily seen that if $U$ and $V$ are relatively thin and $U' \subseteq U$, $V' \subseteq V$, then $U'$ and $V'$ are relatively thin. In particular, every subset of a thin set is thin. One can also check that any finite union of sets, each of which is thin relative to a fixed set $V$, is also thin relative to $V$. Finally, if $K_1,K_2 \subseteq M$ are compact, then $((K_1,K_2))$ is closed in $G$. Hence, if $K_1$ and $K_2$ are compact and relatively thin then $((K_1,K_2))$ is compact.

\begin{definition}[Def. 1.1.2 of \cite{Palais}]
\label{cartan}
$M$ is a {\em Cartan $G$-space} if every $p \in M$ has a thin neighborhood.
\end{definition}

(Note that if $G$ is compact this notion is trivial.) Since every subset of a thin set is thin, if $M$ is a Cartan $G$-space then it has a basis of thin open sets.

\begin{proposition}[Prop. 1.1.4 of \cite{Palais}]
\label{T1}
If $M$ is a Cartan $G$-space, then each orbit of $G$ in $M$ is closed in $M$ (so that $M/G$ is a $T_1$ topological space) and each isotropy group $G_p$ ($p \in M$) is compact.
\end{proposition}
\qcd

\begin{remark}
\label{remark1}
{\em If $M$ is a Cartan $G$-space, $M/G$ may not be Hausdorff even if
%$M$ is a smooth manifold and
$G$ is
%a Lie group
acting
%smoothly and
freely on $M$ (conf. counterexample with $G = (\mathbb{R},+)$ after Theorem 2 of Ref. \cite{Harris}).}
\end{remark}

A simple (and well-known) characterization of when $M/G$ {\em is} Hausdorff is given by the following
\begin{lemma}
\label{interesting}
$M/G$ is Hausdorff iff the following set is closed in $M \times M$:
\[
R = \{(p,gp) \, | \, p \in M, g \in G \}.
\]
\end{lemma}
\qcd

\begin{definition}[Def. 1.2.1 of \cite{Palais}]
\label{smallsets}
A subset $S \subseteq M$ is {\em small} if $\forall p \in M$, there exists a neighborhood $U \ni p$ which is thin relative to $S$.
\end{definition}
One can easily check that every subset of a small set is small, and that a finite union of small sets is small. Moreover, if $S$ is a small subset of $M$ and $K$ is a compact subset of $M$, then $K$ is thin relative to $S$. (In fact, $K$ has a neighborhood which is thin relative to $S$.)

\begin{definition}[Def. 1.2.2 of \cite{Palais}]
\label{properset}
$M$ is a {\em proper $G$-space} if every point $p \in M$ has a small neighborhood.
\end{definition}
It is immediate that if $G$ is compact, then $M$ is automatically a proper $G$-space.

\begin{proposition} [Prop. 1.2.3 of \cite{Palais}]
\label{properimpliescartan}
If $M$ is a proper $G$-space, then it is a Cartan $G$-space.
\end{proposition}
{\em Proof.} Let $p \in M$ and let $U \ni p$ be a small neighborhood. Let $V$ be a neighborhood of $p$ which is thin relative to $U$. Then $U \cap V$ is a thin neighborhood of $p$.
\qcd
We shall see later on that the converse of this result holds iff $M/G$ is Hausdorff (cf. Theorem \ref{propermany}). Nevertheless, the following weaker statement is valid.

\begin{proposition}[Prop. 1.2.4 of \cite{Palais}]
\label{invariant}
If $U \subseteq M$ is a thin open set in $M$, then $GU$ is a proper $G$-space. Hence, if $M$ is a Cartan $G$-space, then every point of $M$ is contained in an invariant open set which is a proper $G$-space.
\end{proposition}

\qcd

The following theorem is a fusion of Thm. 1.2.9 of \cite{Palais} and Prop. 9.13 in Ref. \cite{Lee}, and gives a number of distinct characterizations of a proper $G$-space applicable to our context.
\begin{theorem}
\label{propermany}
If $M$ is a $G$-space, then the following are equivalent.
\begin{itemize}
\item[1)] Given $p,q \in M$, there exist relatively thin neighborhoods $U$ and $V$ of $p$ and $q$, respectively.
\item[2)] $M$ is a Cartan $G$-space and $M/G$ is Hausdorff.
\item[3)] $M$ is a proper $G$-space.
\item[4)] Every compact subset of $M$ is small.
\item[5)] For every $K \subseteq M$ compact, $((K,K))$ is compact in $G$ (thus, every compact subset is thin).
\item[6)] Given any sequences $(p_n)$ in $M$ and $(g_n)$ in $G$ such that both $(p_n)$ and $(g_np_n)$ converge in $M$, some subsequence of $(g_n)$ converges in $G$.
\end{itemize}
\end{theorem}
\qcd

\begin{remark}
\label{rightglitch}
{\em The more standard notion of properness is that $\rho$ a proper action if the map $\tilde{\rho}: (g,p) \in G \times M \mapsto (gp,p) \in M \times M$ is proper in the usual sense that inverse images of compact sets are compact. However, it is very easy to check that $M$ is a proper $G$-space iff $\rho$ is proper in this latter sense. Moreover, we have been using a {\em left} action, but it is clear that all notions of thinness, smallness, etc., above have their analogues for {\em right} actions as well. On the other hand, one can of course define a right action $\rho_{OP}: M \times G \rightarrow M$ associated with $\rho$ by $\rho_{OP}(p,g):= \rho(g^{-1},p)$, $\forall g \in G$, $\forall p \in M$. Now, it is easy to check that subsets $U$ and $V$ of $M$ are relatively thin wrt $(M,\rho)$ iff they are relatively thin wrt $(M, \rho_{OP})$, so $(M, \rho)$ is a Cartan (resp. proper) $G$-space iff $(M,\rho_{OP})$ is Cartan (resp. proper).}
\end{remark}

The next characterization is less trivial.
\begin{theorem}
\label{invariantmetric}
$M$ is a proper $G$-space iff it is Cartan and it admits an invariant Riemannian metric (i.e., a Riemannian metric $h$ for which $\rho_g^{\ast}h =h$, $\forall g \in G$).
\end{theorem}
{\em Proof.} The ``only if'' part is exactly Thm. 4.3.1 of \cite{Palais}, together with Proposition \ref{properimpliescartan}. For the ``if'' part, suppose that $M$ is a differentiable Cartan $G$-space, and let $h$ be an invariant Riemannian metric on $M$. Let $d_h$ denote the corresponding distance function on $M$. It is easily checked that the invariance of $h$ implies that
\[
d_h(gp,gq) = d_h(p,q), \forall p,q \in M, \forall g \in G.
\]
Pick any sequences $(p_n)$ in $M$ and $(g_n)$ in $G$ such that both $(p_n)$ and $(g_np_n)$ converge in $M$, say to $p$ and $q$ respectively. Now,
\[
d_h(g_np,q) \leq d_h(g_np,g_np_n) + d_h(g_np_n, q) \equiv d_h(p,p_n) + d_h(g_np_n, q) \rightarrow 0,
\]
so $g_np \rightarrow q$, and in particular $q \in Gp$, since the latter set is closed (cf. Proposition \ref{T1}). Therefore, $q = gp$, say, so that $g^{-1}g_np \rightarrow p$. Let $U \ni p$ be a thin neighborhood of $p$. Since eventually $g^{-1}g_n \in ((U,U))$, we can, up to passing to a subsequence, assume that $(g^{-1}g_n)$, and hence $(g_n)$, converge in $G$. The result now follows by item (6) of Theorem \ref{propermany}.
\qcd

The following standard result clarifies the importance of having a free and proper action in our context.
\begin{theorem}
\label{lee}
Suppose that $\rho$ is free and proper, then
\begin{itemize}
\item[i)] the orbit space $M/G$ is a topological manifold of dimension $\dim M - \dim G$, and has a unique smooth structure with the property that the projection $\pi: M \rightarrow M/G$ is a smooth submersion, and
\item[ii)] $\pi: M \rightarrow M/G$ defines a smooth principal $G$-bundle over the base manifold $M/G$, with associated right action given by $\rho_{OP}$.
\end{itemize}
In particular, if $G=\mathbb{R}^m$, this bundle is actually trivial, and so, $M$ is diffeomorphic to $\mathbb{R}^m\times M/\mathbb{R}^m$.

Conversely, if $\Pi: M \rightarrow B$ is a smooth principal $G$-bundle over some smooth base manifold $B$ with the associated right action of $G$ on $M$ given by $\rho_{OP}$, then this action (and hence $\rho$) is free and proper, and there exists a diffeomorphism $\phi: B \rightarrow M/G$ such that $\phi \circ \Pi = \pi$.
\end{theorem}
{\em Proof.} See, e.g., in \cite{Lee}, p. 218, Thm. 9.16 for a proof of (i). Finally, see, e.g., \cite{Vandenban}, Sections 12 and 13, for proofs of the other statements, up to the assertion about the case $G=\mathbb{R}^m$ just below (ii), for which one can see, e.g., Theorem 5.7, p. 58 of \cite{KN}.
\qcd

\begin{remark}
\label{nearmanifold}
{\em Suppose that $M$ is a differentiable Cartan $G$-space. Then $M/G$ is $T_1$ (cf. Proposition \ref{T1}) but it is Hausdorff iff $M$ is proper (cf. Theorem \ref{propermany}), and hence $M/G$ may not be a manifold in general. However, $M$ admits a countable basis $\{U_n\}_{n \in \mathbb{N}}$ of thin open sets, and for each $n \in \mathbb{N}$, $GU_n$ is a proper differentiable $G$-space (cf. Proposition \ref{invariant}). Suppose moreover that the action of $G$ is free. Then $\pi(U_n) = U_n/G$ is a smooth manifold of dimension $\dim M - \dim G$. This in turn implies that $M/G$ has a countable basis and is locally homeomorphic to $\mathbb{R}^{\dim M - \dim G}$, but not necessarily Hausdorff, and therefore $M/G$ is what S. Harris has termed a {\em near manifold} in \cite{Harris}. Moreover, in this case $\pi: GU_n \rightarrow U_n/G$ is a principal $G$-bundle for each $n \in \mathbb{N}$, so $\pi: M \rightarrow M/G$ is still a (topological) principal $G$-bundle.}
\end{remark}

The last proposition of this section establishes, for an important special case, a significant means to ensure the properness of the action. Although it is a rather simple result, we were unable to find it elsewhere with this generality, and thus we have included its proof here.
\begin{proposition}
\label{contractible}
Suppose that $G = \mathbb{R}^m$ ($m < n$) (with its additive group structure), and that it acts freely on $M$. Then the following assertions are equivalent.
\begin{itemize}
\item[i)] $M$ is a proper $G$-space.
\item[ii)] There exists a codimension $m$ submanifold $\Sigma \subset M$ with the following property: for each $p \in M$, there exists a unique $x \in \Sigma$ in the orbit of $p$ by $G$. In other words, each orbit intersects $\Sigma$ exactly once. (Such submanifold is called a {\em slice} for the action.)
\end{itemize}
If one (hence both) of these assertions holds, $\Sigma$ is properly embedded and diffeomorphic to $M/\mathbb{R}^m$.
\end{proposition}
{\em Proof.} ($(i) \Rightarrow (ii)$) \\
If $(i)$ holds, then by Theorem \ref{lee}, we have that $M$ is a trivial smooth principal bundle over $M/\mathbb{R}^m$ with structure group $(\mathbb{R}^m,+)$, and so,
%But then by a standard result (see, e.g., Theorem 5.7, p. 58 of \cite{KN}) this bundle is actually trivial, so
we can pick a trivialization $\tau: M \rightarrow \mathbb{R}^m \times S$, where $S:= M/\mathbb{R}^m$.
Define $\Sigma := \tau ^{-1} (0 \times S)$. Since $\tau$ is in particular a diffeomorphism, the inverse image of the properly embedded ($n-m$)-dimensional submanifold $0 \times S$ of $\mathbb{R}^m \times S$ is also a properly embedded submanifold of $M$ of the same dimension.

Because of the principal bundle structure, $\tau$ can be chosen so that the following diagrams commute:

\[
\begin{array}{lll} \xymatrix{M \ar[d]_-{\rho_g} \ar[r]^-{\tau} & \mathbb{R}^m \times S \ar[d]^-{t_g \times \mathbb{I}} \\
M\ar[r]_-{\tau} & \mathbb{R}^m \times S } & & \xymatrix{M \ar[rr]^-{\tau} \ar[dr]_-{\pi} & & \mathbb{R}^m \times S \ar[dl]^-{proj_2} \\ & S},

\end{array} \]
for all $g \in \mathbb{R}^m$. Here, $proj_2$ denotes the projection onto the second cartesian factor, and $t_g$ is the translation on $\mathbb{R}^m$ by the element $g \in \mathbb{R}^m$ (since $G$ is Abelian, the distinction between right and left actions is irrelevant). Given $p \in M$, write $\tau(p) = (g,x) \in \mathbb{R}^m \times S$, and let $q := \tau^{-1}(0,x)$. Then
\[
\tau(p) = (g,x) = (t_g \times \mathbb{I}) \cdot (0,x) = (t_g \times \mathbb{I}) \circ \tau(q) = \tau (\rho_g (q)),
\]
and hence $p = gq$, i.e, $q \in \Sigma \cap Gp$. Let $q' \in \Sigma \cap Gp$. Then $p = g' q'$ for some $g' \in \mathbb{R}^m$ and $\tau(q') = (0,y)$ for some $y \in S$. Thus,
\[
(0,x)= \tau(q) = \tau ((g'-g)q') = (t_{g'-g} \times \mathbb{I})\tau(q') = (g'-g,y),
\]
and hence $x=y$ and $g=g'$, and the orbit through $p$ intersects $\Sigma$ only once. \\

($(ii) \Rightarrow (i)$) \\
 Let $\Sigma \subset M$ satisfy the condition in $(ii)$. Let $\Phi:= \rho|_{\mathbb{R}^m \times \Sigma}: \mathbb{R}^m \times \Sigma \rightarrow M$. Because of condition $(ii)$ and since the action is free, $\Phi$ is a smooth bijection, and hence a homeomorphism by Invariance of Domain.

Now, let $(p_n) \subseteq M$ and $(g_n) \subseteq \mathbb{R}^m$ be sequences such that $p_n \rightarrow p$ and $g_np_n \rightarrow q$. We can then write $p = \Phi(t,x)$ and $q=\Phi(s,y)$ for uniquely defined $t,s \in \mathbb{R}^m$ and $x,y \in \Sigma$. Likewise, for each $n \in \mathbb{N}$, $p_n = \Phi(t_n, x_n)$. Since $\Phi$ is a homeomorphism, $g_n+ t_n \rightarrow s$, $t_n \rightarrow t$, and $x_n \rightarrow x,y$, so $x=y$ and $(g_n)$ converges to $g:=s-t$, and the action is proper by item (6) of \ref{propermany}. This concludes the equivalence.

To show the final assertion of the theorem, we can now assume that both $(i)$ and $(ii)$ are valid. Condition $(ii)$ means that $\pi |_{\Sigma}: \Sigma \rightarrow M/\mathbb{R}^m$ is a smooth bijection, and since $\pi: M \rightarrow M/\mathbb{R}^m$ is a smooth submersion by $(i)$ (cf. Theorem \ref{lee}), $d \pi _p$ defines an isomorphism between $T_p\Sigma$ and $T_{\pi(p)}(M/\mathbb{R}^m)$ for each $p \in \Sigma$. By the Inverse Function Theorem, $\Sigma$ and $M/\mathbb{R}^m$ are thus diffeomorphic. The properness of the embedding follows easily by using the fact that $\Phi$ is a homeomorphism.
\qcd

\begin{corollary}
\label{corcontractible}
Under the assumptions and notation in Proposition \ref{contractible}, let $\Sigma \subset M$ be a slice. If $S \subseteq M$ is a submanifold invariant by the $\mathbb{R}^m$-action on $M$, i.e., $gS \subseteq S$ for all $g \in \mathbb{R}^m$, and $\Sigma$ is transversal to $S$, then the induced action on $S$ is proper.
\end{corollary}
{\em Proof.} Just note that the induced action is well-defined, remains free and $\Sigma \cap S$ will be a slice for it.
\qcd

%%%%%%%%%%%%%%%%%%%%%%%%%%%%%%%%%%%%%%%%%%%%%%%%%%%%%%%%%%%%%
\section{The Splitting Problem I: Topological Splitting and Causality}\label{0.3}
%%%%%%%%%%%%%%%%%%%%%%%%%%%%%%%%%%%%%%%%%%%%%%%%%%%%%%%%%%%%%

Here and hereafter, we shall fix a smooth Lorentzian metric\footnote{Up to this point, we were using `$g$' for elements of the group $G$. From now on, however, in order to avoid confusion with the traditional notation for metrics, and since we will presently specialize to $G= \mathbb{R}$ anyway, we will use $t,s$, etc. for elements of the group.} $g$ on $M$. Since we shall assume the existence of an (everywhere non-zero) causal vector field $X$ on $M$, there will be no real loss of generality for our main results if we assume that $(M^n,g)$ is a {\em spacetime}, i.e, that $M$ is connected, $n \geq 2$ and $(M,g)$ is time-oriented, and we will always do so in what follows.

In this section we will focus on the interplay between causal conditions on $(M,g)$ and the existence of a {\em topological} splitting in the presence of a suitable group action, deferring to the next section the examination of the existence of concomitant metric splittings. We shall take $G= \mathbb{R}$ with its standard addition operation throughout this section, so that our action $\rho$ may be taken to be the flow of a fixed smooth complete, everywhere non-zero vector field $X: M \rightarrow TM$. Due to its greater applicability in the Lorentzian Geometry context, in what follows we shall be interested only in the case where $X$ is causal.

Physically reasonable causality restrictions on $(M,g)$ arise also naturally from a mathematically viewpoint due to the following consideration. Since we wish to give $M$ a (trivial) principal $\mathbb{R}$-bundle structure, we shall need $\rho$ to be free and proper. In particular, since $X$ cannot have closed orbits, $(M,g)$ is required to be at least {\it chronological} (i.e., must have no closed timelike curves) if $X$ is timelike, or {\it causal} (i.e., without closed causal curves) if $X$ is causal. In addition, in order to ensure properness, we shall often impose stronger causality assumptions.

For vector fields we have the following nice refinemenet of Theorem \ref{invariantmetric}.
\begin{corollary}
\label{completemetric}
If the action $\rho$ generated by $X$ is proper, then there exists a {\em geodesically complete} Riemannian metric $g_0$ on $M$ for which $X$ is a Killing vector field.
\end{corollary}
{\em Proof.} Pick any complete Riemannian metric $g_c$ on $M$. By Proposition \ref{contractible}, there exists a smooth, properly embedded (hence closed) hypersurface $\Sigma \subset M$ which each orbit of $X$ intersects exactly once. Thus, the map $P: M \rightarrow \Sigma$ given by associating each $p \in M$ with the intersection of its orbit with $\Sigma$ is surjective retraction which leaves $\Sigma$ invariant pointwise. Indeed, using arguments similar to those in the proof of Proposition \ref{contractible}, we see that $P$ is smooth. Now, since $\Sigma$ is closed, the induced metric $g_c|_{\Sigma}$ is complete by Hopf-Rinow theorem, and since $X$ is complete, the pullback metric $g_0:= P^{\ast} (g_c|_{\Sigma})$ is complete and invariant by the flow of $X$ by construction.
\qcd

We start the next part by adapting the definition of an {\em ancestral pair} as given in \cite{HL}.

\begin{definition}
\label{ancestral}
Suppose $X$ is future-directed causal. A pair $(p,q) \in M \times M$ is an {\em ancestral pair} (resp. {\em weakly ancestral pair}) {\em for $\rho$ in $(M,g)$} if $\mathbb{R}p \subseteq I^{-}(q)$ (resp. $\mathbb{R}p \subseteq \overline{I^{-}(q)}$).
\end{definition}

\begin{example}
\label{basicexample}
{\em Let $M = \mathbb{R}^2 \setminus \{(0,0)\}$ with the flat metric $ds^2 = -dt^2 + dx^2$, where we take $(t,x)$ to be the standard cartesian coordinates, and time orientation such that $\partial _t$ is future-directed. Let $\lambda \in C^{\infty}(M)$ be any positive function for which the future-directed null vector field
\[
X = \lambda(\partial_t - \partial_x)
\]
is complete\footnote{Recall that any everywhere non-zero vector field on a manifold can be rescaled as a complete vector field.}. The generated action $\rho$ has both, ancestral pairs (e.g., $((-1,1),(1,0))$) and weakly ancestral pairs (e.g., $((-1,1),(1,-1))$).}
\end{example}
\begin{remark}
\label{aboutancestral}
{\em It is immediate that if there are no weakly ancestral pairs for $\rho$ in $(M,g)$, then there are no ancestral pairs either. But the converse fails, as the following simple example shows. Let $M = \{(t,x) \,|\, t,x \in \mathbb{R}\} \setminus \{(0,x) \,| \, x \geq 0 \}$, with the flat metric given by $ds^2 = -dt^2 + dx^2$, and time orientation such that $\partial _t$ is future-directed. Let $\lambda \in C^{\infty}(M)$ be any positive function for which the future-directed null vector field
\[
X = \lambda(\partial_t - \partial_x)
\]
is complete. Then any pair of the form $((-a,a),(b,-b))$ for $a,b>0$ is weakly ancestral for the generated action $\rho$, but there are no ancestral pairs for $\rho$.}
\end{remark}
Recall that $(M,g)$ is {\em distinguishing} if $\forall p,q \in M$,
\[
I^{+}(p) = I^{+}(q) \mbox{ or }  I^{-}(p) = I^{-}(q) \Rightarrow p=q.
\]
It is well-known \cite{BE} that this condition implies that $(M,g)$ is causal, but not the converse. $(M,g)$ is {\em causally continuous} if it is distinguishing and the set-valued functions $I^{\pm}: p \in M \mapsto I^{\pm}(p) \in \mathbb{P}(M)$\footnote{Here, $\mathbb{P}(M)$ denotes the power set of $M$.} are {\em outer continuous} in the following sense: for any $p \in M$ and any compact set $K \subseteq M \setminus \overline{I^{\pm}(p)}$, there exists a neighborhood $U \ni p$ such that $K \subseteq M \setminus \overline{I^{\pm}(p')}$ whenever $p' \in U$. Again, this condition implies that $(M,g)$ is stably causal \cite{BE}, but not the converse. (The definition of stable causality is recalled below, just before Proposition \ref{immediate}.) Indeed, it is easy to check that the spacetime in the Remark \ref{aboutancestral} is not causally continuous, but  Proposition 3.43, p. 89 of \cite{BE} implies that it is stably causal.  The following proposition shows that no similar construction can be made for causally continuous spacetimes.

\begin{proposition}
\label{causalcontinuity}
Suppose $X$ is future-directed causal and $\rho$ its generated action. If $(M,g)$ is causally continuous and there are no ancestral pairs, then there are no weakly ancestral pairs for $\rho$ in $(M,g)$.
\end{proposition}
{\em Proof.} It suffices to show that $\forall p,q \in M$,
\[
p \in I^{-}(q) \Rightarrow \overline{I^{-}(p)} \subseteq I^{-}(q).
\]
Thus, let $p \in I^{-}(q)$ and  $r \in \overline{I^{-}(p)}\setminus I^{-}(q)$. In particular, $r \in \partial I^{-}(p)\cap \partial I^{-}(q)$. Causal continuity then implies that $p,q \in \partial I^{+}(r)$ (cf. Lemma 3.42 of \cite{MS}), which is impossible since $\partial I^{+}(r)$ is achronal.
\qcd

There is a natural physical interpretation, in the context of Mathematical Relativity, for the non-existence of ancestral pairs for $\rho$ when $X$ is a future-directed causal vector field. Namely, it may be interpreted as a  form of ``cosmic censorship'': if $(p,q)$ is ancestral for $\rho$, then the whole ``history'' represented by the orbit of $p$ is ``visible'' from $q$, which in turn can be interpreted as saying that ``$q$ gets information from infinity'' (since each orbit is complete). Such a situation is excluded, for instance, if the so-called strong cosmic censorship (which is equivalent to global hyperbolicity) holds (see \cite{Penrose} for a detailed discussion). Accordingly, we have the following
\begin{proposition}
\label{noancestral}
If $(M,g)$ is globally hyperbolic and $X$ is future-directed causal, then $\rho$ has no (weakly) ancestral pairs in $(M,g)$.
\end{proposition}
{\em Proof.} Let $p,q$ be any two points in $M$. If $(p,q)$ were weakly ancestral pair, then the future-directed, future-inextendible causal curve $ t \in [0, +\infty) \mapsto tp \in M$ would be entirely contained in the compact set $J^{+}(p) \cap \overline{I^{-}(q)} \equiv  J^{+}(p) \cap J^{-}(q)$, which is impossible due to the strong causality of $(M,g)$.
\qcd

\begin{remark}
\label{quickone}
{\em Proposition \ref{noancestral} implies that any complete causal vector field in widely studied spacetimes such as Minkowski, deSitter, Schwarzschild, Schwarschild-Kruskal, Schwarzchild-deSitter, and many Robertson-Walker cosmological models have no (weakly) ancestral pairs. However, any of the standard causal conditions other than global hyperbolicity is insufficient {\em per se} to ensure the non-existence of ancestral pairs (see the paragraph below Corollary \ref{goodcausal}). On the other hand, it is not the case that causality conditions play an exclusive role in the existence (or not) of (weakly) ancestral pairs. The presence of these pairs can be precluded by additional natural conditions on $(M,g)$, even under rather mild causality assumptions. One such important situation in which there are no ancestral pairs is when $X$ is a future-directed timelike {\em conformal Killing} vector field (i.e., $L_Xg = fg$, for some $f \in C^{\infty}(M)$), a case which includes important non-globally hyperbolic spacetimes such as Anti-deSitter.}
\end{remark}
\begin{proposition}
\label{conformastationary}
If $(M,g)$ is chronological and $X$ is a future-directed timelike conformal Killing vector field, then $\rho$ has no ancestral pairs in $(M,g)$.
\end{proposition}
{\em Proof.} Given $p,q \in M$, Corollary 3.2 of Ref. \cite{HL} shows in particular that the orbit of $p$ will eventually enter $I^{+}(q)$ and, since $(M,g)$ is chronological, it must therefore leave $I^{-}(q)$, i.e., $(p,q)$ is not ancestral.
\qcd

Theorem \ref{harris} and Corollary \ref{harris2} below offer an alternative proof of the main result in \cite{Harris} (cf. Theorem 1 and Corollary 1 of that reference), using a different but equivalent phrasing more suited to the formalism we have adopted here. We also include in the same results an alternative version of Theorem 1.2 of \cite{HL}.
\begin{theorem}
\label{harris}
 If $(M,g)$ is chronological and $X$ is future-directed timelike, then $\rho$ is free and $M$ is a Cartan $\mathbb{R}$-space. If in addition $\rho$ has no ancestral pairs in $(M,g)$, then $M$ is a proper $\mathbb{R}$-space, and so, it is diffeomorphic to $\mathbb{R}\times M/\mathbb{R}$.
\end{theorem}
{\em Proof.} The free character of the action is in any case immediate from chronology.

For the first part, let $p \in M$. If $p$ does not have a thin neighborhood, then there exist sequences $(t_n) \subseteq \mathbb{R}$ and $(p_n) \subseteq M$ such that $p_n,t_np_n \rightarrow p$ and $|t_n| \rightarrow +\infty$. Up to passing to a subsequence, we may then assume that $t_n \rightarrow +\infty$, the case if $t_n \rightarrow -\infty$ being analogous. Pick any positive number $t_0$. Since $p << t_0p$ and the chronological order is an open relation, there exist open neighborhoods $U \ni t_0p$ and $V \ni p$ such that $p' << q'$ whenever $p' \in V$ and $q' \in U$. For large enough $n$, $t_n >t_0$, $t_np_n \in V$ and $t_0p_n \in U$, so
\[
t_0p_n << t_np_n << t_0p_n,
\]
contradicting chronology.

Assume now that $\rho$ has no ancestral pairs in $(M,g)$. We only need to show that $M/\mathbb{R}$ is Hausdorff (cf. item (2) of Theorem \ref{propermany}). Suppose, by way of contradiction, that this is false. Then, by Lemma \ref{interesting}, there exist sequences $(t_n) \subseteq \mathbb{R}$ and $(p_n) \subseteq M$ such that $p_n \rightarrow p$ and $t_np_n \rightarrow q$, while $q \notin \mathbb{R}p$. Clearly $(t_n)$ cannot be bounded, so we may assume, up to passing to a subsequence, that $|t_n| \rightarrow +\infty$, and indeed that $t_n \rightarrow +\infty$, for otherwise we may work with $(-t_n)$ just as well, performing straightforward adaptations. Let $t, \epsilon >0$. Since $tp << (t+\epsilon)p$, there exist open sets $U_p \ni tp$ and $V_p \ni (t+ \epsilon)p$ such that $p' << p''$ whenever $p' \in U_p$ and $p'' \in V_p$. Similarly, there exist open sets $U_q \ni (-\epsilon)q$ and $V_q \ni q$ such that $q' << q''$ whenever $q' \in U_q$ and $q'' \in V_q$. Eventually, $t_n>t+2\epsilon$, $(t+ \epsilon)p_n \in V_p$ and $(t_n - \epsilon)p_n \in U_q$, so that
\[
tp <<  (t+ \epsilon)p_n << (t_n - \epsilon)p_n << q,
\]
and we conclude that $(p,q)$ is an ancestral pair, contrary to our assumption.
\qcd
\begin{corollary}
\label{harris2}
If $(M,g)$ is chronological and $X$ is a future-directed timelike conformal Killing vector field, then $M$ is a proper $\mathbb{R}$-space.
\end{corollary}
{\em Proof.} Immediate from Theorem \ref{harris} and Proposition \ref{conformastationary}.
\qcd

\begin{remark}
\label{general}
{\em Corollary \ref{harris2} admits an alternative proof which does not mention ancestral pairs as follows. Having established that the action is Cartan, the strategy is now to apply Theorem \ref{invariantmetric}, so one must prove that $M$ admits an invariant Riemannian metric. Now, note that Lemma 2.1 of \cite{JS} implies (and in any case it is very easy to check) that if $X$ is a conformal Killing vector field for $g$, then it is a Killing vector field for the conformally rescaled metric $\overline{g} = -g/g(X,X)$. Since $(M,\overline{g})$ is still chronological, we may drop the overline and just assume that $X$ is Killing. But then $X$ is also Killing for the Riemannian metric given by
\[
g_R(v,v) = g(v,v) - \frac{2}{g(X,X)}g(X,v)^2,
\]
which is thus invariant. This argument is used, in a slightly different way, by Harris in \cite{Harris}.}
\end{remark}

We now turn our attention to the case when $X$ is causal. We first note that this case might be reduced to previous results under some rather favorable circumstances. We consider one such case. Denote by $Lor(M)$ the set of Lorentzian metrics on $M$. Recall that we can introduce a partial order in $Lor(M)$ as follows: for $g_1,g_2 \in Lor(M)$,
\[
g_1 < g_2 \Leftrightarrow \forall v \in TM\setminus \{0\}, \mbox{ if }g_1(v,v)\leq 0, \mbox{ then } g_2(v,v) <0.
\]
Recall also that $(M,g)$ is {\em stably causal} iff there exists some $g' \in Lor(M)$, such that $g < g'$ and that $(M,g')$ is causal.
\begin{proposition}
\label{immediate}
Assume that $X$ is future-directed causal for $(M,g)$, and that there exists some $g' \in Lor(M)$ such that:
\begin{itemize}
\item[i)] $g < g'$ with $(M,g')$ causal (so, in particular, $(M,g)$ is stably causal), and
\item[ii)] $X$ is a conformal Killing vector field for $(M,g')$.
\end{itemize}
Then $\rho$ is proper and free.
\end{proposition}
{\em Proof.} Since $g<g'$, $X$ is future-directed timelike in $(M,g')$, and the result now follows by Corollary \ref{harris2} applied to $(M,g')$.
\qcd

Even in the absence of such special conditions, there is a direct analogue of Theorem \ref{harris} for causal $X$, which however requires stronger hypotheses.
\begin{theorem}
\label{strongcausality}
If $(M,g)$ is strongly causal and $X$ is a future-directed causal vector field, then $\rho$ is free and $M$ is a Cartan $\mathbb{R}$-space. If in addition $\rho$ has no weakly ancestral pairs in $(M,g)$, then $M$ is a proper $\mathbb{R}$-space, and so, it is diffeomorphic to $\mathbb{R}\times M/\mathbb{R}$.
\end{theorem}
{\em Proof.} That $\rho$ is free is immediate from causality. Suppose that $M$ is not Cartan. Then, for some $p \in M$ there exist sequences $(p_n) \subseteq M$ and $(t_n) \subseteq \mathbb{R}$ such that $p_n, t_np_n \rightarrow p$, but $|t_n| \rightarrow +\infty$. Up to passing to a subsequence, we can assume that $t_n \rightarrow +\infty$, the case if $t_n \rightarrow -\infty$ being analogous.

Fix $U \in p$ a relatively compact neighborhood of $p$. From strong causality there exists an open neighborhood $V\subseteq U$ of $p$ for which any causal curve segment with endpoints in $V$ is entirely contained in $U$. Again, up to passing to a subsequence, we can assume that $p_n,t_np_n \in V$ for all $n$.

Now, fix a complete Riemannian metric $h$ on $M$. For each $n \in \mathbb{N}$, let $\tilde{\alpha}_n : t\in [0, +\infty) \mapsto tp_n \in M$ be the orbit of $p_n$, and $\alpha_n:[0, +\infty) \rightarrow M$ be its $h$-arc length reparametrization. For a unique $s_n \in [0,+ \infty)$, $\alpha_n(s_n) = t_np_n$, and we have $\alpha_n[0,s_n] \subset U$. Since $X$ is future-directed, these are all future-directed causal curves; and therefore, by the Limit Curve Lemma, we may pick a future directed, causal limit curve $\alpha:[0, +\infty) \rightarrow M$ with $\alpha(0) = p$ and such that $\alpha_n |_C \rightarrow \alpha|_C$ $h$-uniformly in compact subsets $C \subseteq [0, +\infty)$. Clearly, $s_n \rightarrow +\infty$, so given any $s>0$, eventually $s_n >s$, in which case $\alpha_n[0,s] \subseteq U$. But then, from the uniform convergence in $[0,s]$, $\alpha[0,s] \subseteq \overline{U}$, and therefore we conclude that $\alpha[0,+\infty)$ is entirely contained in the compact subset $\overline{U}$, contradicting (a well known consequence of) strong causality.

The rest of the proof proceeds much like the second part of the proof of Theorem \ref{harris}. Assume that $\rho$ has no weakly ancestral pairs in $(M,g)$. It again suffices to show that $M/\mathbb{R}$ is Hausdorff. If not, by Lemma \ref{interesting}, there exist sequences $(t_n) \subseteq \mathbb{R}$ and $(p_n) \subseteq M$ such that $p_n \rightarrow p$ and $t_np_n \rightarrow q$, while $q \notin \mathbb{R}p$. Clearly $(t_n)$ cannot be bounded, so we may assume, up to passing to a subsequence, that $|t_n| \rightarrow +\infty$, and again without loss of generality that $t_n \rightarrow +\infty$. Let $t>0$, and fix any $r \in I^{+}(q)$. For large enough $n$ we then have
\[
tp_n < t_np_n << r,
\]
and hence $tp \in \overline{I^{-}(r)}$. Since $t$ is arbitrary, we conclude that $(p,r)$ is a weakly ancestral pair, contrary to our assumption.
\qcd
\begin{remark}
\label{whyancestral}
{\em In Theorems \ref{harris} and \ref{strongcausality}, we needed, in addition to the causality conditions, to require the non-existence of ancestral pairs and weakly ancestral pairs, respectively. To see that extra conditions are often necessary to obtain properness, consider again the example in the Remark \ref{aboutancestral}. Recall that the spacetime therein is stably causal (and hence strongly causal), but (say) $p=(-1,1)$ and $q =(1,-1)$ form a weakly ancestral pair and have disjoint orbits. Given any two neighborhoods $U \ni p$ and $V \ni q$ there exists an orbit of $\rho$ which crosses both $U$ and $V$. We conclude that the orbit space $M/\mathbb{R}$ is not Hausdorff, so $\rho$ cannot be proper, although it is still Cartan. As pointed out in the cited remark, in this example there are no ancestral pairs, and hence it also illustrates the fact that the condition of non-existence of weakly ancestral points in Theorem \ref{strongcausality} cannot be weakened to non-existence of ancestral pairs.}
\end{remark}
The following direct consequence of Theorem \ref{strongcausality} and Proposition \ref{noancestral} shows that if the causality condition is strong enough, then the condition of non-existence of (weakly) ancestral pairs is unnecessary.
\begin{corollary}
\label{GH}
If $(M,g)$ is globally hyperbolic and $X$ is future-directed causal, then $\rho$ is free and proper.
\end{corollary}
\qcd
Corollary \ref{GH} also admits an alternative proof. If $(M,g)$ is globally hyperbolic, it will admit an acausal smooth Cauchy hypersurface \cite{BS}; if we use this fact, then this corollary is also an immediate consequence of Proposition \ref{contractible}.

The requirement that there are no weakly ancestral pairs can be weakened to non-existence of ancestral pairs if $(M,g)$ is causally continuous:
\begin{corollary}
\label{goodcausal}
If $(M,g)$ is causally continuous, $X$ is future-directed causal, and $\rho$ has no ancestral pairs in $(M,g)$, then $M$ is a proper $\mathbb{R}$-space.
\end{corollary}
{\em Proof.} Immediate from Proposition \ref{causalcontinuity} and Theorem \ref{strongcausality}.
\qcd

The condition that there are no (weakly) ancestral pairs for $\rho$ in $(M,g)$, although sufficient and rather natural, is {\em not} necessary to ensure that $\rho$ is a proper, free action. To see this, consider the following construction, adapted from an example given by Harris and Low in \cite{HL} (cf. p. 30 of that reference, just after the proof of Theorem 1.2 therein). Let $M$ be the strip $\{(t,x) \in \mathbb{R}^2 \,|\, 2x <t<2x+1 \}$ and let $g$ be the flat metric given by $ds^2 = -dt^2+dx^2$, with time-orientation such that $\partial_t$ is future-directed. Let $X_1 = \lambda_1 \partial_t$ and $X_2 = \lambda_2 (\partial_t + \partial_x)$ be vector fields defined on $M$, where $\lambda_i \in C^{\infty}(M)$ ($i=1,2$) is any positive function such that $X_i$ is complete. We have ancestral pairs for the actions of $X_1$ (say $((1/2,0),(3/2,1))$) and $X_2$ (say $((-1/2,-1/2),(1/2,0))$), but in both cases, $M/\mathbb{R} = \mathbb{R}$. It easy to check that the underlying spacetime $(M,g)$ is causal and all past an future causal cones $J^{\pm}(p)$ are closed, and thus it is {\em causally simple}, a causal condition which implies causal continuity, and is weaker only than global hyperbolicity in the causal ladder \cite{MS}. In particular, $(M,g)$ is strongly causal, so the action of both $X_i$'s is Cartan, and hence proper and free. (Of course, it is easy to see that $(M,g)$ is not globally hyperbolic.)

The previous example, although instructive, is somewhat contrived. As we shall see in the following result (which is analogous to Corollary \ref{harris2}), a more natural important situation in which the non-existence of weakly ancestral pairs can be dispensed with (but retaining appropriate causality assumptions) is when $X$ is a {\em conformal Killing causal} vector field.
\begin{theorem}
\label{jose}
If $(M,g)$ is strongly causal and $X$ is a conformal Killing causal complete vector field, then $M$ is a proper $\mathbb{R}$-space, and so, it is diffeomorphic to $\mathbb{R}\times M/\mathbb{R}$.
\end{theorem}
{\em Proof.} From Theorem \ref{strongcausality}, $M$ is a Cartan ${\mathbb R}$-space (and the ${\mathbb R}$ action associated to $X$ is free). According to Theorem \ref{propermany}, in order to prove that $M$ is also a
proper ${\mathbb R}$-space, it suffices to show that $M/{\mathbb R}$ is Hausdorff.
So, assume by contradiction that it is not, and denote by $\pi:M\rightarrow M/{\mathbb R}$ the corresponding bundle projection.
%Let $g_0$ be some Riemannian metric on $M$ and $d$ its associated distance.
There exist two points $p, q\in M$ such that all neighborhoods of $\pi(p)$ and $\pi(q)$ intersect. So, there exists $X$ orbits $\gamma_{n}$ and points $p_n, q_n\in \gamma_n$ such that $p_n\rightarrow p$, $q_n\rightarrow q$. Denote by $\gamma_p$ and $\gamma_q$ the $X$ orbits which pass through $p$ and $q$, resp. It is not a restriction to assume that, at least, $\gamma_q$ is null at some point (and so, it is necessarily a null geodesic).
In fact, otherwise, the open set $O$ determined by the timelike $X$ orbits of $M$ is a strongly causal (thus, chronological) spacetime with a conformal Killing timelike complete vector field $X\mid_O$ such that $O/\mathbb{R}$ is not Hausdorff, and so, $O$ is not a proper $\mathbb{R}$-space, in contradiction with Corollary \ref{harris2}.

Let $\Sigma_p$ and $\Sigma_q$ be some small hypersurfaces passing through $p$ and $q$, resp., which are crossed at most once by any integral curve of $X$. We can suppose without restriction that $p_n\in\Sigma_p$ and $q_n\in \Sigma_q$ for all $n$, and the existence of a sequence $\{t_n\}\subset{\mathbb R}$ such that
$\gamma_n(0)=p_n$, $\gamma_n(t_n)=q_n$.
Since $\pi(p)\neq\pi(q)$, necessarily $t_n\rightarrow\infty$ as $n\rightarrow\infty$.

Next, consider $\Sigma_p\setminus \{p\}$ endowed with {\em generalized spherical} coordinates $(r,\theta_1,\ldots,\theta_m)$, where $r\in (0,\infty)$, $\theta_1\in (-\pi,\pi)$, $\theta_{i}\in (-\pi/2,\pi/2)$, $i=2,\ldots,m$, with $m:={\rm dim}(M)-2$. By readjusting $\theta_i$ we can suppose that (up to a convenient subsequence of $p_n$, and thus, of $q_n$) $\theta_i(p_n)=0$ for all $i$ and all $n$\footnote{In fact,
up to a subsequence, we can assume that the radial component $r_n$ of $p_n$ is
strictly decreasing. Then, we can connect $p_n$ between them by a smooth
curve inside $\Sigma_p\setminus\{p\}$ with strictly decreasing radial component $r$. By
defining $\theta_i=0$ for all $i$ on this curve, it is easy to globally define $\theta_i$ on
$\Sigma_p\setminus\{p\}$ for all $i$ with the required property.}. Consider the neighborhood $U$ around $\gamma_p$ generated by acting the flow $\phi$ of $X$ on $\Sigma_p$. Note that $U\setminus\gamma_p$ can be parametrized by coordinates ($t$ the parameter of the flow $\phi$)
\[
(t,r,\theta_1,\theta_2,\ldots,\theta_{m})\in{\mathbb R}\times{\mathbb R}^+\times (-\pi,\pi)\times (-\pi/2,\pi/2)\times\cdots\times (-\pi/2,\pi/2).
\]
%Then, $U$ is parameterized by $(t,r,\theta)$, being $t$ the parameter of the inegral curves of $X$.
%, being $t$ the parameter of the flow $\phi$ of $Y$.
So, we can write $\gamma_n(0)=(0,r_n,0,\ldots,0)$, $\gamma_n(t_n)=(t_n,r_n,0,\ldots,0)$ for all $n$.

Denote by $T:D\subset{\mathbb R}^+\rightarrow {\mathbb R}$ the smooth function determined by $(T(r),r,0,\ldots,0)\in\Sigma_q$ ($D$ is an open subset of ${\mathbb R}^+$ containing $r_n$ for all $n$). In particular, $T(r_n)=t_n\rightarrow\infty$ as $n\rightarrow\infty$. We can also assume that $T'(r_n)\rightarrow -\infty$ as $n\rightarrow\infty$ (in fact, otherwise, argue with $q_n$ replaced by
$(T(\bar{r}_n),\bar{r}_n,0,\ldots,0)$, and thus, $p_n$ replaced by $(0,\bar{r}_n,0,\ldots,0)$, where $\{\overline{r}_n\}$ is some sequence with $\bar{r}_n\rightarrow 0$ and
$T'(\bar{r}_n)\rightarrow\infty$). Then, consider the new coordinate
\[
s(t,r):=\int_0^t\lambda(\tau,r)d\tau,
\]
where $\lambda$ is the positive function given by
\[
\lambda(t,r):=\exp(-\eta(r)t(T(r)-t)),
\]
and $\eta(r)$ is some positive function such that
\[
\begin{array}{c}
s(T(r),r)\left(=\int_0^{T(r)}\lambda(\tau,r)d\tau\right)=\int_0^{T(r)}\exp(-\eta(r)\tau(T(r))-\tau)d\tau=1 \\ \hbox{(since $T'(r_n)<0$ for $n$ big enough, necessarily $\eta'(r_n)<0$ for $n$ big enough)}.
\end{array}
\]
%Then,
%\[
%s(t(r),r)\left(=\int_0^{t(r)}\lambda(\tau,r)d\tau\right)=\int_0^{t(r)}\exp(-a(r)\tau(t(r))-\tau)d\tau=1\quad\hbox{for all $r$.}
%\]
In particular, $\partial_t=\lambda\partial_s$. Moreover, the following properties hold:
\begin{equation}\label{gg}
\begin{array}{c}
\lambda(t_n,r_n)=\exp(-\eta(r_n)t_n(t_n-t_n))=1,\quad \lambda(0,r_n)=\exp(-\eta(r_n)0(t_n-0))=1,
\\  \\
\frac{\partial}{\partial r}\mid_{r=r_n}\lambda(t,r)=-(\eta'(r_n)t(t_n-t)+\eta(r_n)t(T'(r_n)-t))\exp(-\eta(r_n)t(t_n-t))>0\;\; \forall t\in (0,t_n)
\\ \\
\frac{\partial}{\partial r}\mid_{r=r_n}\left(\int_0^1\frac{ds}{\lambda(s,r)}\right)=\frac{\partial}{\partial r}\mid_{r=r_n}t(1,r)=T'(r_n)\rightarrow -\infty\quad\hbox{as $n\rightarrow\infty$.}
\end{array}
\end{equation}
%\begin{equation}\label{gg}
%\begin{array}{c}
%\lambda(t_n,r_n)=\lambda(0,r_n)=1\quad\hbox{for all $n$},
%\\ \\ s_n=s(t_n,r_n)=1\quad\hbox{for all $n$,} \\ \\
%%s(t_n,r_n)\;\;\hbox{bounded}\quad\hbox{and}\quad
%\frac{\partial}{\partial r}\mid_{r=r_n}\left(\int_0^{1}\frac{ds}{\lambda(s,r)}\right)\rightarrow\infty\quad\hbox{as $n\rightarrow\infty$}.
%\end{array}
%\end{equation}
On the other hand, there exist $v_p,w^i_p\in T_pM$, with $\{X_p,v_p,w^1_p,\ldots,w^{m}_p\}$ linearly independent\footnote{In order to justify the linearly independence of these vectors, we can assume that our coordinates $(s,r,\theta_1,\ldots,\theta_m)$ takes the neighborhood $U\setminus \gamma_p$
into a cylinder of ${\mathbb R}^{m}$ with the central axis removed. Consider the metric
$h$ on $U\setminus \gamma_p$ induced by the usual one in the cylinder via the
coordinates $(s,r,\theta_1,\ldots,\theta_m)$. Clearly, the coordinate vectors $X, \partial_r,
\partial_{\theta_1},\ldots,\partial_{\theta_m}$ are orthogonal on $U$ with this metric. Assume for
instance that the limit vectors $v_p$, $w^1_p$ of $\partial_r$, $\partial_{\theta_1}$,
resp., are collinear at $p$. Then, the metric $h$ cannot be continuously
extended to a metric $\bar{h}$ on $U$ (since, otherwise, the orthogonal
vectors $v_p$, $w^1_p$ would be collinear, an absurd). But this is false because such an extension $\bar{h}$ can be constructed by taking the metric on U
induced by the usual one in the full cylinder via the {\em cartesian}
coordinates.}, such that the following limits hold, up to a subsequence,
\[
\partial_s\mid_{\gamma_n(0)}\rightarrow X_p,\qquad \partial_r\mid_{\gamma_n(0)}\rightarrow v_p,\qquad  \frac{1}{r_n}\partial_{\theta_i}\mid_{\gamma_n(0)}\rightarrow w_p^i\quad\hbox{for all $i$}.
\]
Analogously, there exist $v_q,w^i_q\in T_qM$, with $\{X_q,v_q,w^1_q,\ldots,w^{m}_q\}$ linearly independent, such that the following limits hold, up to a subsequence,
\[
\partial_s\mid_{\gamma_n(1)}\rightarrow X_q,\qquad \partial_r\mid_{\gamma_n(1)}\rightarrow v_q,\qquad  \frac{1}{r_n}\partial_{\theta_i}\mid_{\gamma_n(1)}\rightarrow w^i_q\quad\hbox{for all $i$}.
\]
Even more, by slightly moving the points $p_n$ on $\Sigma_p$ (and thus, $q_n$ on $\Sigma_q$) conveniently, we can additionally assume that
\[
a(p)=g(X_p,v_p)\neq 0,\qquad f^i(p)=g(X_p,w^i_p)\neq 0\quad\hbox{$i=1,\ldots, m$,}
\]
and thus,
\begin{equation}\label{r}
a(\gamma_n(0))\neq 0,\quad f^i(\gamma_n(0))\neq 0\quad\hbox{$i=1,\ldots, m$,}\quad\hbox{for all $n$ big enough.}
\end{equation}
In general, the metric $g$ can be written in local coordinates $(s,r,\theta_1,\ldots,\theta_{m})$ as
\[
g=\alpha ds^2+2adsdr+bdr^2+\sum_i d\theta_i(c^id\theta_i+2e^idr+2f^ids)+2\sum_{i\neq j}h^{ij}d\theta_id\theta_j.
\]
Then, a direct computation shows that the condition of being $X(=\partial_t)=\lambda \partial_s$ a conformal Killing vector field in $(M,g)$ is equivalent to:
\begin{equation}\label{k}
\begin{array}{c}
\lambda\alpha_s + 2\alpha\lambda_s=\beta\alpha
\qquad
\lambda b_s+2a\lambda_r=\beta b,\qquad \lambda c^i_s=\beta c^i,\qquad \lambda e^i_s+f^i\lambda_r=\beta e^i,\qquad h^{ij}_s=0 \\ \\ (\lambda a)_s + \alpha\lambda_r=\beta a,\qquad (\lambda f^i)_s=\beta f^i,\qquad\beta\in C^{\infty}(M).
\end{array}
\end{equation}
From the sixth equation in (\ref{k}), and taking into account (\ref{r}) and the second line in (\ref{gg}), we deduce that the metric function $a$ does not vanish on $\gamma_n([0,t_n))$. Similarly, but now using the seventh equation in (\ref{k}) and again (\ref{r}),
 we also deduce that the metric functions $\{f^i\}_{i=1}^m$ do not vanish all along $\gamma_n([0,t_n))$.
So, manipulating the second and sixth equations, and the forth and seventh ones,\footnote{Note that this manipulation requires to divide the corresponding equations by $a$ and $f^i$, resp.; so, this is why we have previously ensured that these functions do not vanish.} we deduce
\[
\left(\frac{b}{a\lambda}\right)_s=\left(2-\frac{\alpha b}{a^2}\right)\left(\frac{1}{\lambda}\right)_r\geq 2\left(\frac{1}{\lambda}\right)_r,\qquad \left(\frac{e^i}{f^i\lambda}\right)_s=\left(\frac{1}{\lambda}\right)_r\quad i=1,\ldots,m.
\]
By integrating these equations along $\gamma_n([0,1])$, and taking into account the limit in (\ref{gg}), we obtain
\begin{equation}\label{d1}
\begin{array}{l}
\frac{b(\gamma_n(1))}{a(\gamma_n(1))\lambda(\gamma_n(1))}-\frac{b(\gamma_n(0))}{a(\gamma_n(0))\lambda(\gamma_n(0))}\geq 2\int_0^1\left(\frac{1}{\lambda(s,r_n)}\right)_rds=2\frac{\partial}{\partial r}\mid_{r=r_n}\left(\int_0^1\frac{ds}{\lambda(s,r)}\right)\rightarrow\infty,
\\ \\
\frac{e^i(\gamma_n(1))}{f^i(\gamma_n(1))\lambda(\gamma_n(1))}-\frac{e^i(\gamma_n(0))}{f^i(\gamma_n(0))\lambda(\gamma_n(0))}=\int_0^1\left(\frac{1}{\lambda(s,r_n)}\right)_rds=\frac{\partial}{\partial r}\mid_{r=r_n}\left(\int_0^1\frac{ds}{\lambda(s,r)}\right)\rightarrow\infty\quad i=1,\ldots,m.
\end{array}
\end{equation}
%So, from the inequality in (\ref{gg}),
%\begin{equation}\label{d1}
%|b(\gamma_n(\overline{s}_n))-b(\gamma_n(s_n))|=2|\mu(r_n,\theta_n)|\int_{t_n}^{\overline{t}_n}\frac{dt}{\lambda(r_n,t)}\geq 2|\mu(r_n,\theta_n)|(\overline{t}_n-t_n).
%\end{equation}
But
\[
\frac{b(\gamma_n(0))}{a(\gamma_n(0))\lambda(\gamma_n(0))}\rightarrow \frac{g(v_p,v_p)}{g(X_p,v_p)\lambda(p)},\qquad b(\gamma_n(1))\rightarrow g(v_q,v_{q}),\qquad \lambda(\gamma_n(1))\rightarrow \lambda(q),
\]
hence
\[
\lim_n a(\gamma_n(1))=0.
\]
Analogously, since
\[
\frac{e^i(\gamma_n(0))}{f^i(\gamma_n(0))\lambda(\gamma_n(0))}\rightarrow \frac{g(v_p,w^i_p)}{g(X_p,w^i_p)\lambda(p)},\qquad e^i(\gamma_n(1))/r_n\rightarrow g(v_q,w^i_q)\quad i=1,\ldots,m,
\]
necessarily
\[
\lim_n\frac{f^i(\gamma_n(1))}{r_n}=0\quad i=1,\ldots,m.
\]
In concusion,
\[
\begin{array}{l}
g(X_q,v_q)=g(\lim_n\partial_s\mid_{\gamma_n(1)},\lim_n\partial_r\mid_{\gamma_n(1)})=\lim_n g(\partial_s\mid_{\gamma_n(1)},\partial_r\mid_{\gamma_n(1)})=\lim_n a(\gamma_n(1))=0,
\\
g(X_q,w^i_q)=g(\lim_n\partial_s\mid_{\gamma_n(1)},\lim_n\frac{1}{r_n}\partial_{\theta_i}\mid_{\gamma_n(1)})=\lim_n g(\partial_s\mid_{\gamma_n(1)},\frac{1}{r_n}\partial_{\theta_i}\mid_{\gamma_n(1)})=\lim_n\frac{f^i(\gamma_n(1))}{r_n}=0\;\;\forall i.
\end{array}
\]
So, we have deduced that the null vector $X_q$ is orthogonal to $v_p$ and $w^i_p$ for all $i$, and so, $X_q$ belongs to the radical of $g_q$, an absurd. \qcd

\noindent {\bf Note.}
The proof that $M$ is a
proper $\mathbb{R}$-space admits the following more direct argument when dim$(M)=2$. According to Theorem \ref{invariantmetric}, it suffices to prove the existence of some Riemannian metric $g_R$ on $M$ such that $X$ is also Killing for $g_R$. Consider the {\em unique} vector field $Y$ on $M$ such that $Y_p$ is null and $g(X_p,Y_p)=-1$ for every $p\in M$. Then, the following properties hold:
\[
g(X,X)\equiv 0,\quad g(Y,Y)\equiv 0,\quad g(X,Y)\equiv -1.
\]
Next, consider the unique Riemannian metric $g_R$ on $M$ such that
\[
g_R(X,X)\equiv 1,\quad g_R(Y,Y)\equiv 1,\quad g_R(X,Y)\equiv 0.
\]
If we denote by $\phi$ the flow associated to the Killing vector field $X$, the stages $\phi_t$ are isometries for $g$. Moreover, $Y$ is $g$-metrically determined by $X$. Therefore,
\begin{equation}\label{eq1}
\phi_{t}^{*}(X_p)=X_{\phi_{t}(p)}\quad\hbox{and}\quad \phi_{t}^{*}(Y_p)=Y_{\phi_{t}(p)}.
\end{equation}
So, according to (\ref{eq1}), we have
\[
\begin{array}{c}
g_R(\phi_t^*(X_p),\phi_t^*(X_p))=g_R(X_{\phi_t(p)},X_{\phi_t(p)})=1=g_R(X_{p},X_{p}) \\ \\
g_R(\phi_t^*(Y_p),\phi_t^*(Y_p))=g_R(Y_{\phi_t(p)},Y_{\phi_t(p)})=1=g_R(Y_{p},Y_{p}) \\ \\
g_R(\phi_t^*(X_p),\phi_t^*(Y_p))=g_R(X_{\phi_t(p)},Y_{\phi_t(p)})=0=g_R(X_{p},Y_{p}).
\end{array}
\]
In conclusion, we deduce that the stages $\phi_t$ are isometries for $g_R$, and thus, $X$ is Killing for $g_R$.

\begin{remark}
\label{perplexing}
{\em We do not know at present if, for causal or strongly causal $(M,g)$, and in analogy to what occurs with timelike vector fields (cf. Proposition \ref{conformastationary}), the fact that $X$ is conformal Killing and causal implies that $\rho$ has no weakly ancestral points, but we conjecture that this is the case.}
\end{remark}

%%%%%%%%%%%%%%%%%%%%%%%%%%%%%%%%%%%%%%%%%%%%%%%%%%%%%%%%%%%%%
\section{Metric Splitting of Stationary Spacetimes: the Javaloyes-S\'{a}nchez Splitting Theorem} \label{0.4}
%%%%%%%%%%%%%%%%%%%%%%%%%%%%%%%%%%%%%%%%%%%%%%%%%%%%%%%%%%%%%

In the previous section, a number of topological splitting results have been obtained which greatly expand the scope of the timelike case studied by Harris {\em et al.} to include the causal case. The next natural step is to investigate the consequences of these results for the global geometric structure of $(M,g)$ when a complete causal Killing vector field $X$ is defined thereon. However, it turns out that in order to obtain more definite results, one must fix the causal character of $X$ to be either timelike everywhere or null everywhere. As we mentioned before, the timelike case is well understood \cite{Harris,JS}. In order to motivate our approach, it is instructive to review the (well-known) basic features of this case. This is the purpose of this short section.

Let $X$ be a complete timelike Killing vector field on $(M,g)$, generating an isometric $\mathbb{R}$-action $\rho$ on $M$.  Recall (cf. Proposition \ref{contractible}) that a {\em slice} for a free $\mathbb{R}$-action on a manifold is a hypersurface $\Sigma \subset M$ which intersects each orbit exactly once, and that such a slice exists if and only if $\rho$ is proper. Note that a standard stationary spacetime (\ref{stationarymetric}) is certainly chronological - indeed causally continuous, see Corollary 3.2 of \cite{JS} - and $t$ is a time function, so any hypersurface $t=const.$ is an {\em achronal} (hence acausal) spacelike slice. Thus, the action of $X=\partial_t$ is free and proper by Proposition \ref{contractible}.

Assume that $M$ is a principal (trivial) $\mathbb{R}$-bundle over the quotient $S:= M/\mathbb{R}$ with bundle projection $\pi:M \rightarrow S$. We have seen that the latter condition occurs \cite{Harris} when $(M,g)$ is chronological (cf. also Corollary \ref{harris2}). Since $X$ is Killing, the orthogonal distribution $X^{\perp} \subset TM$ is covariant, in the sense that $(d\rho_t)_p(X^{\perp}_p) = X^{\perp}_{\rho_t(p)}$ for all $t \in \mathbb{R}$ and all $p \in M$, and of course $T_pM = X^{\perp}_p \oplus \mathbb{R} X_p$. In other words, $X^{\perp}$ is a principal (Ehresmann) $\mathbb{R}$-connection on $\pi: M \rightarrow S$, for which the orthogonal complements of $X$ at each point are the horizontal spaces, and having an associated $\rho$-invariant 1-form $\theta \in \mathfrak{X}^{\ast}(M)$. Indeed, $\theta$ is uniquely defined by the conditions $\theta(X) \equiv 1$ and $\theta(X^{\perp})  \equiv 0$. Thus, the ``horizontal'' part of its curvature $d \theta$ measures the integrability of the orthogonal distribution, i.e., $d\theta|_{X^{\perp} \wedge X^{\perp}} =0$ iff $X^{\perp}$ is integrable. In this case, $(M,g)$ is said to be {\em static}.

Now, for each $p \in M$, $\ker d\pi_p$ is the span of $X_p$, and since $\pi$ is a submersion $d \pi_p$ has maximal rank, i.e., $\dim X_p^{\perp} = \dim T_{\pi(p)}S$, so $d\pi_p|_{X_p^{\perp}}: X_p^{\perp} \rightarrow T_{\pi(p)}S$ is an isomorphism. Since $X$ is Killing, the induced inner product on $T_{\pi(p)}S$ is the same for all points along the orbit of $p$, so there exists a unique Riemannian metric $g_S$ on $S$ defined by requiring that this pointwise isomorphisms are linear isometries. (Hence, when $S$ is endowed with this metric, $\pi$ becomes a semi-Riemannian submersion.) The splitting of vectors into vertical and horizontal parts implies that the metric $g$ can then be written as
\begin{equation}
\label{splittingtimelike}
g = -\beta ^2 \theta \otimes \theta + \pi^{\ast}g_S,
\end{equation}
where $\beta := \sqrt{|g(X,X)|}$. Since $\beta$ is $\rho$-invariant, it can be regarded as a function $\beta_S \in C^{\infty}(S)$. Given any trivialization $\mathcal{T}: M \rightarrow \mathbb{R} \times S$ of the $\mathbb{R}$-bundle, there exists a unique one-form $\omega_{S}$ on $S$ for which $(\mathcal{T}^{-1})^{\ast} \theta = dT + proj_2^{\ast} \omega_{S}$, where $proj_2: \mathbb{R} \times S \rightarrow S$ is the standard projection onto the second Cartesian factor, and $T$ is the projection onto the first factor. We can then regard $\mathcal{T}$ as an isometry between $(M,g)$ and the product space $(\mathbb{R} \times S, \tilde{g})$ endowed with the metric
 \begin{equation}
\label{splittingtimelike2}
\tilde{g} = -\beta_S ^2 (dT + proj_2^{\ast} \omega_{S}) \otimes (dT + proj_2^{\ast} \omega_{S}) + proj_2^{\ast} g_S.
\end{equation}
By this isometry, $X$ is taken to $\partial_T$.

Eq.(\ref{splittingtimelike2}) does describe a metric splitting, but it has the following defect as far as applications in Physics go. Despite the appearances, $T$ {\em need not} be a time function for the spacetime $(\mathbb{R} \times S, \tilde{g})$ unless $\omega_S \equiv 0$, and it is then static. Indeed, the level sets of any smooth time function are acausal hypersurfaces. Here, however, these are essentially copies of $S$, and the induced ``metric'' on $S$ is
\[
g_S^{ind} = g_S -\beta_S ^2 \omega_{S} \otimes \omega_{S},
\]
which need not have a definite signature, and so it need not be a (semi-)Riemannian metric. On the other hand, it might well be, say, negative (semi)definite along regular curves in $S$, violating acausality.

A natural question is whether we can choose the trivialization $\mathcal{T}$ in such a way that $g_S^{ind}$ {\em is} positive definite, in which case $(\mathbb{R} \times S, \tilde{g})$ will be a standard stationary spacetime. As pointed out in the Introduction, it has been shown in \cite{JS} that this can be done if and only if $(M,g)$ is distinguishing. The gist of the argument in \cite{JS} will reappear in a somewhat different form in the null case we treat below, so we partly review it here. But for convenience of our presentation, we shall do so in a slightly different form.
\begin{proposition}[Javaloyes-S\'{a}nchez \cite{JS}]
\label{JavaloyesMiguel}
Let $\rho$ be a isometric $\mathbb{R}$-action on a chronological spacetime $(M,g)$ generated by the complete timelike Killing vector field $X$. Then
\begin{itemize}
\item[i)] $\rho$ admits a spacelike slice if and only if $(M,g)$ is distinguishing.
\item[ii)] $(M,g)$ is isometric to a standard stationary spacetime (with $X$ identified with $\partial_t$) if and only if $\rho$ admits a spacelike slice.
\end{itemize}
\end{proposition}
{\em Proof.} $(i)$ \\
Using the notation above, note that if $g_S^{ind}$ is positive-definite, then $T$ {\em is} a time function (its gradient is actually timelike - cf. Corollary 3.2 of \cite{JS}), and in particular $(\mathbb{R} \times S, \tilde{g})$ (and hence $(M,g)$) is stably causal, and therefore also distinguishing. Conversely, if $(M,g)$ is a distinguishing stationary spacetime, then it is causally continuous (Proposition 3.1 of \cite{JS}); in other words, the distinguishing and continuous causality properties are equivalent for stationary spacetimes. In particular, $(M,g)$ is stably causal, and hence there exists \cite{BS} a {\em temporal function}, i.e., a smooth function $f:M \rightarrow \mathbb{R}$ with timelike gradient $\nabla f$. A given level set $\Sigma$ of $f$ is of course an acausal spacelike hypersurface, and thus the integral curves of $X$ intersect $\Sigma$ at most once. The key to the proof of the main theorem in \cite{JS} is showing that any orbit of $X$ does indeed intersect $\Sigma$, so that $\Sigma$ is a slice. The authors of \cite{JS} accomplish this by an argument involving the local standard stationary form (\ref{stationarymetric}) of the metric. The details of this part will not be relevant to us, so we omit them here. \\
$(ii)$\\
The ``only if'' part is immediate from Proposition \ref{contractible}. For the converse, let $\Sigma \subset M$ be a spacelike slice, so that in particular $\rho$ is free and proper, and $M$ is a principal trivial $\mathbb{R}$-bundle over the quotient $S= M/\mathbb{R}$ with bundle projection $\pi:M \rightarrow S$. Denote the inclusion of $\Sigma$ by $i_{\Sigma} : \Sigma \hookrightarrow M$. Just as in the proof of Proposition \ref{contractible}, we can check that $\pi|_{\Sigma}:\Sigma \rightarrow S$ will be a diffeomorphism. Hence, it is straightforward to check that the bijection $\mathcal{T} : M \rightarrow \mathbb{R} \times S$ such that
\[
\mathcal{T}^{-1} = \rho \circ (Id_{\mathbb{R}} \times i_{\Sigma}) \circ (Id_{\mathbb{R}} \times (\pi|_{\Sigma})^{-1})
\]
is indeed a bundle trivialization for which $g_S^{ind}$ is positive definite.
\qcd
\begin{remark}
\label{similar}
{\em For later reference, we emphasize that Proposition \ref{JavaloyesMiguel} divides the Javaloyes-S\'{a}nchez splitting theorem in two parts:
\begin{itemize}
\item[1)] A {\em causal condition}, giving a characterization of the existence of a spacelike slice via a causal condition on the spacetime, and
\item[2)] A {\em geometric condition on the orbits}, giving a characterization of standard stationary spacetimes as precisely those stationary spacetimes for which the action admits a such a spacelike slice.
\end{itemize}
}
\end{remark}
%%%%%%%%%%%%%%%%%%%%%%%%%%%%%%%%%%%%%%%%%%%%%%%%%%%%%%%%%%%%%
\section{The Splitting Problem II: Spacetimes with a Parallel Null Vector Field and the Global Structure of Brinkmann spacetimes}\label{0.5}
%%%%%%%%%%%%%%%%%%%%%%%%%%%%%%%%%%%%%%%%%%%%%%%%%%%%%%%%%%%%%

Most of the above constructions in the previous section go awry when $X$ is a {\em null} (complete) Killing vector field. Again, if the action $\rho$ generated by $X$ is free and proper (e.g., if $(M,g)$ is strongly causal - cf. Theorem \ref{jose}), then we still have that $M$ is a principal (trivial) $\mathbb{R}$-bundle over the quotient $S:= M/\mathbb{R}$ with bundle projection $\pi:M \rightarrow S$. The orthogonal distribution $X^{\perp}$ is still well-defined and it is indeed covariant in the above sense, but it fails to give a pointwise direct sum splitting of the tangent space into a vertical and horizontal parts, because each $X_p^{\perp}$ is now degenerate and hence does {\em not} define a principal connection as before. Since the fibers of $X^{\perp}$ are degenerate $(n-1)$-dimensional subspaces of the tangent spaces, it gives rather a filtration $\mathbb{R}X \subset X^{\perp} \subset TM$ of the tangent bundle, where $\mathbb{R}X$ denotes the line bundle defined by the directions of $X$. Moreover, an attempt to use the procedure outlined above to define a metric on $S$ also fails, as the resulting tensor $g_S$ is now degenerate\footnote{However, a simple metric splitting similar to the timelike one {\em can} be performed if $(M,g)$ is a globally hyperbolic spacetime \cite[Prop. 2.2]{BCFl}.}.

Now, it has long been known from the study of null submanifolds of semi-Riemannian manifolds \cite{duggal} that one must often consider {\em non-canonical} choices of extra geometric structures thereon in order to give the ``horizontal'' or ``transversal'' information we need. This can indeed be done in special cases, as we will see in more detail below. We start with a fairly general definition.

\begin{definition}
\label{conjugate}
Let $X \in \Gamma(TM)$.
\begin{itemize}
\item[i)] A vector field $Y \in \Gamma(TM)$ is said to be {\em conjugate} to $X$ if $g(X,Y) =1$ and $[X,Y] =0$.
\item[ii)] A 1-form $\omega \in \Omega ^1(M)$ is {\em conjugate} to $X$ if $\omega(X) =1$ and $L_X \omega =0$.
\end{itemize}
\end{definition}

The notions of conjugate vector fields and 1-forms are related as follows.
\begin{proposition}
\label{onetoone}
Let $X \in \Gamma(TM)$ be a Killing vector field on the spacetime $(M,g)$\footnote{This result can be obviously generalized to any semi-Riemannian manifold.}. Then the standard map $Y \in \Gamma(TM) \mapsto \omega_{Y}:=g(Y, \, . \, ) \in \Omega ^1(M)$ induces a one-to-one correspondence between vector fields on $M$ conjugate to $X$ and 1-forms on $M$ conjugate to $X$.
\end{proposition}
{\em Proof.} Just note that under the assumption that $X$ is Killing we have, for $Y \in \Gamma(TM)$,
\[
L_X \omega_Y (Z) = g([X,Y],Z)
\]
for every $Z \in \Gamma(TM)$. From this, the result easily follows.
\qcd

\begin{remark}
{\em Let $X$ be any smooth vector field. The set of 1-forms conjugate to $X$ is naturally an affine space with respect to the vector space
\[
\{ \omega \in \Omega ^1(M) \, | \, L_X \omega =0 \mbox{ and } \omega(X) =0\}.
\]
In particular, if $X$ is null Killing and $\omega$ is a conjugate 1-form, then so is $\omega + f \chi$, where $f \in C^{\infty}(M)$ is $X$-invariant, (i.e., $X(f) =0$), and $\chi = g(X, \, . \,)$ is the 1-form metrically associated with $X$. More generally, note that the definition above does not require either completeness of the vector field $X \in \Gamma(TM)$ or that it be Killing or have any fixed causal character. However, if $X$ is a complete vector field generating a free and proper action, so that $M$ is a principal $\mathbb{R}$-bundle over the orbit space $M/\mathbb{R}$, then it is clear that the connection 1-forms on $M$ are exactly the 1-forms conjugate to $X$, which are thus in one-to-one association with Ehresmann connection on the underlying bundle.}
\end{remark}

If $X$ is a smooth complete timelike Killing vector field on the chronological spacetime $(M,g)$, the question of existence and multiplicity of such conjugate forms is overshadowed by the fact that the 1-form $\theta$ associated with the orthogonal distribution is by far the most natural choice. (Its associated conjugate vector field is $X/g(X,X)$.) However, for null Killing vector fields it becomes important, and can be characterized as follows.

\begin{proposition}
\label{conjugateexistence}
Let $X \in \Gamma(TM)$ be a null Killing vector field on $(M,g)$ (not necessarily complete).
\begin{itemize}
\item[a)] There exists a conjugate vector field $Y \in \Gamma(TM)$ (and hence a conjugate 1-form) if and only if there exists a Riemannian metric $h$ on $M$ for which $X$ is also a Killing vector field.
\item[b)] If $Y \in \Gamma(TM)$ is conjugate to $X$, then the vector field $\hat{Y}:= Y - \frac{1}{2} g(Y,Y)X$ is a {\em null} vector field also conjugate to $X$, and in particular the two-dimensional subspace $span \{X_p,Y_p\} = span\{X_p,\hat{Y}_p\} \subset T_pM$ ($p \in M$) is timelike.
\end{itemize}
\end{proposition}
{\em Proof.} We will only develop the proof of (a), since (b) is trivial.

For the ``only if'' part, let $Y \in \Gamma(TM)$ be conjugate to $X$. Note that since $g(X,Y)=1$, $Y$ is everywhere non-parallel to $X$. We can define a Riemannian metric $h$ on $M$ as follows. Put $h(X,X) = h(Y,Y) =1$, $h(X,Y) =0$, and for any $V,W$ $g$-orthogonal to both $X$ and $Y$ (and hence necessarily spacelike or zero), set $h(X,V) = h(Y,V) = 0$ and $h(V,W) = g(V,W)$, extending by linearity.

For each $p \in M$, we introduce the notation ${\cal H} \subseteq TM$ for the distribution of timelike 2-planes given by ${\cal H}_p := \mbox{span }\{X_p,Y_p\}$, $\forall p \in M$, and denote by ${\cal H}^{\perp}$ its associated $g$-orthogonal distribution. Since each fiber of ${\cal H}$ is timelike, we of course have the splitting $TM = {\cal H} \oplus {\cal H}^{\perp}$. (In particular, all the fibers of ${\cal H}^{\perp}$ are spacelike.) Then we can write
\begin{equation}
\label{commutator}
[X,Y] = -g(Y,[X,Y])X - g(X,[X,Y])Y + [X,Y]^{\perp},
\end{equation}
where $[X,Y]^{\perp}$ denotes the component of $[X,Y]$ along ${\cal H}^{\perp}$. Now, since $X$ is Killing, we have
\[
0= (L_Xg)(X,Y) = X(g(X,Y)) - g([X,X],Y) - g(X,[X,Y]) \equiv  -g(X,[X,Y]),
\]
and
\[
0= (L_Xg)(Y,Y) = X(g(Y,Y)) - 2g(Y,[X,Y]) \equiv  - 2g(Y,[X,Y]),
\]
whence we conclude that $[X,Y] = [X,Y]^{\perp}$.

We now show that $X$ is Killing with respect to $h$. Computing the Lie derivatives of the metric $h$, we get first
\begin{eqnarray}
(L_Xh)(X,X) &=& X(h(X,X)) - 2 h([X,X],X) \equiv 0, \\
(L_Xh)(X,Y) &=& X(h(X,Y)) - h([X,X],Y) - h(X,[X,Y]) \equiv 0,\\
(L_Xh)(Y,Y) &=& X(h(Y,Y)) - h([X,Y],Y) - h(Y,[X,Y]) \equiv 0,
\end{eqnarray}
where each term on the right hand sides vanishes either from the definition of $h$ or from the fact that $[X,Y]$ is $g$-orthogonal (and hence $h$-orthogonal) to both $X$ and $Y$. Now, given any vector field $V$ in ${\cal H}^{\perp}$, we can write
\begin{equation}
\label{commutator2}
[X,V] = -g(Y,[X,V])X - g(X,[X,V])Y + [X,V]^{\perp},
\end{equation}
where as before $[X,V]^{\perp}$ denotes the component of $[X,V]$ along ${\cal H}^{\perp}$, so that
\begin{eqnarray}
(L_Xh)(X,V) &=& X(h(X,V)) - h([X,X],V) - h(X,[X,V]) \\ \nonumber
           &=& -g(Y,[X,V]) \equiv -g(Y,[X,V]) - g([X,Y],V) + X(g(Y,V)) + g([X,Y],V)\\ \nonumber
           &=& (L_Xg)(Y,V) + g([X,Y],V)\equiv g([X,Y],V) \equiv 0,
\end{eqnarray}
where the last equality holds since $X$ is Killing with respect to the background Lorentzian metric $g$. We similarly have, using (\ref{commutator2}) again,
\begin{eqnarray}
(L_Xh)(Y,V) &=& X(h(Y,V)) - h([X,Y],V) - h(Y,[X,V]) \\ \nonumber
           &=& -g(X,[X,V]) -g([X,Y],V) \\ \nonumber
           &=& -g(X,[X,V]) - g([X,X],V) + X(g(X,V)) -g([X,Y],V)\\ \nonumber
           &=& (L_Xg)(X,V) -g([X,Y],V)\equiv -g([X,Y],V) \equiv 0.
\end{eqnarray}
Finally, given $V,W$ in ${\cal H}^{\perp}$,
\begin{eqnarray}
(L_Xh)(V,W) &=& X(h(V,W)) - h([X,V],W) - h(V,[X,W]) \\ \nonumber
           &=& X(g(V,W)) - h([X,V]^{\perp},W) - h(V,[X,W]^{\perp}) \\ \nonumber
           &=& X(g(V,W)) - g([X,V]^{\perp},W) - g(V,[X,W]^{\perp}) \\ \nonumber
           &=& X(g(V,W)) - g([X,V],W) - g(V,[X,W]) \\ \nonumber
           &=& (L_Xg)(V,W) \equiv 0.
\end{eqnarray}
Hence, $X$ is Killing for the metric $h$.

Conversely, to show the ``if'' part, let $h$ is {\em any} Riemannian metric on $M$ for which $X$ is Killing, define a one-form on $M$ by
\[
\omega(Z):= h\left(\frac{X}{h(X,X)},Z\right)
\]
for all $Z \in \Gamma(TM)$. (Note that $X$ is everywhere non-zero!) Clearly, $\omega(X) =1$ and a straightforward computation shows that $X$ being Killing for $h$ translates into $L_X\omega =0$, so $\omega$ is a conjugate 1-form. By Proposition \ref{onetoone}, it defines a unique conjugate vector field $Y$.
\qcd

The following Corollary is an immediate consequence of the previous proposition, Theorem \ref{invariantmetric} and Theorem \ref{jose}.
\begin{corollary}
Assume that $X \in \Gamma(TM)$ is a complete null Killing vector field on the spacetime $(M,g)$, whose flow $\rho$ is a Cartan action on $M$. Then there exists a vector field $Y \in \Gamma(TM)$ conjugate to $X$ if and only if $\rho$ is proper. In particular, if $(M,g)$ is strongly causal then $X$ has a conjugate vector field $Y$, which can be assumed to be null.
\end{corollary}

The following metric splitting result is the analogue, for a null Killing vector field, of (\ref{splittingtimelike}).

\begin{proposition}
\label{nulllemma}
Let $X \in \Gamma(TM)$ be a null vector field on $(M,g)$. Let $Y\in \Gamma(TM)$ be a null conjugate vector field. Then
\begin{equation}
\label{splittingnull}
g = \omega \otimes \chi + \chi \otimes \omega + \hat{g},
\end{equation}
where $\omega = g(Y, \,. \,)$, $\chi  = g(X, \,. \,)$, and $\hat{g}$ is a symmetric $(0,2)$ smooth tensor field over $M$ such that $\forall p \in M$, $\hat{g}_p$ is positive semidefinite and its radical is $span\{X_p,Y_p\}$. In particular, $\hat{g}_p(v,v) >0$ for any non-zero vector $v$ in the rank $n-2$ distribution $X^{\perp} \cap Y^{\perp}$. Moreover, if $X$ is Killing, then $L_X \omega =0$ and $L_X \hat{g} =0$.
\end{proposition}
{\em Proof.} This particular splitting follows immediately from the decomposition
\[
TM = \mathbb{R}X \oplus \mathbb{R}Y \oplus ( X^{\perp} \cap Y^{\perp}),
\]
while the second part follows easily when $X$ is Killing (cf. Proposition \ref{onetoone}) and thus $\omega$ is conjugate in this case.
\qcd

Now, it becomes natural to wonder if the metric splitting (\ref{splittingnull}) corresponds to an underlying topological splitting. The particular form of (\ref{splittingnull}), however, suggests that this putative splitting should be of the form $M \simeq \mathbb{R}^2 \times Q$, instead of the $M \simeq \mathbb{R} \times S$ form we have been considering. This naive hope is misplaced as it stands, as the following simple example shows.
\begin{example}
\label{noncomplete}
{\em Let $(N,\eta)$ be any compact Riemannian manifold, and take $M =\mathbb{R} \times S^1 \times N$ with the metric given by $g = 2dtd \phi +  \eta$. Take $X= \partial_t$ and $Y = \partial_{\phi}$. These are complete, conjugate null vector fields, and $X$ is Killing.}
\end{example}
The fact that $X$ is not gradient in previous example (its metrically associated $1$-form is $\chi = d \phi$, which is not exact) is essential to violate the $M \simeq \mathbb{R}^2 \times Q$ form. In fact, it turns out that this splitting {\em can} be accomplished when $X$ is a null Killing {\em gradient} vector field (with some extra assumptions - see below), i.e., if there exists $u \in C^{\infty}(M)$ with $X = \nabla u$. But this is just saying that $X$ is parallel, i.e., $\nabla X =0$.

\begin{lemma}
\label{simple}
If a Killing vector field $V \in \Gamma(TM)$ of the spacetime $(M,g)$ is gradient, then it is parallel. If $M$ is simply connected, then the converse also holds.
\end{lemma}
{\em Proof.} By the Koszul formula, using the fact that $V$ is Killing we get, $\forall Z,W \in \Gamma(TM)$,
 \[
 2g(\nabla _Z V, W) = g(\nabla_Z V, W) - g(\nabla _W V , Z) = (\mbox{curl}\, V) (Z,W) = 0,
 \]
where the last equality holds since $V$ is gradient. Therefore, $\nabla V =0$. Conversely, if $V$ is parallel, its metrically dual 1-form $\psi = g(V, \, . \,)$ is closed, and if $M$ is simply connected, $\psi$ is also exact, and therefore $V$ is gradient.
\qcd

Lorentzian manifolds endowed with a complete parallel null vector field are often called {\em Brinkmann spaces} after \cite{brinkmann}, where its local structure is described (cf. Eq. (\ref{localBrink})). We shall follow this usage here. Brinkmann spaces have been much studied in the mathematical and physical literature (see, e.g \cite{baum,batat,brinkmann,galaev,leistner,EK,minguzzi1,minguzzi2,walker} and references therein). One minor aspect of such manifolds is that they are always time-orientable \cite{baum}, so fixing a time-orientation (as we have been doing when considering a spacetime) poses no real restriction in this context.

The extra motivation for adopting a null Killing $X = \nabla u$ is that in physical applications the level surfaces $u = const.$ are often interpreted as describing a ``propagating gravitational wave-front at the speed of light'' \cite{BE,EK,wald}.

The primary example of a Brinkmann space is the {\em standard Brinkmann spacetime}, defined below. This is actually a three-parameter family of spacetimes including the so-called {\em $pp$-waves}, which are of great importance in General Relativity, in particular because they are exact vacuum solutions of the Einstein field equation describing the propagation of an (idealized) gravitational wave \cite{BE,EK}.

\begin{definition}
\label{BEspacetimes}
A {\em standard Brinkmann spacetime} is a Lorentzian manifold $(\tilde{M}, \tilde{g})$, where $\tilde{M}=\mathbb{R}^2 \times Q$ for some $n-2$-dimensional smooth manifold $Q$, and
\begin{equation}
\label{reference}
\tilde{g} = dU \otimes (dV + H dU + \Omega) + (dV + H dU + \Omega)\otimes dU + \gamma
\end{equation}
where
\begin{itemize}
\item[i)] $V,U: \mathbb{R}^2 \times Q  \rightarrow \mathbb{R}$ are the standard projections onto the first and second slots of $\mathbb{R}^2$, respectively, and can therefore be interpreted as the standard coordinate functions on $
\mathbb{R}^2$,
\item[ii)] $\gamma$ is a smooth $(0,2)$-tensor on $\mathbb{R}^2 \times Q$, $\Omega$ is a smooth 1-form on $\mathbb{R}^2 \times Q$,
\end{itemize}
and are such that
\begin{itemize}
\item[a)] the radical of $\gamma$ at each $\tilde{p} = (V_0,U_0,x_0) \in \mathbb{R}^2\times Q$ is $span \{\partial_V|_{\tilde{p}},\partial_U|_{\tilde{p}}\}$, so that $x \in Q\mapsto \gamma_{(V_0,U_0,x)}|_{T_xQ \times T_xQ}$ defines a smooth Riemannian metric on $Q$,
\item[b)] $\Omega(\partial_V) = \Omega(\partial_U) =0$, so that at each $\tilde{p} = (V_0,U_0,x_0) \in \mathbb{R}^2\times Q$, $x \in Q\mapsto \Omega_{(V_0,U_0,x)}|_{T_xQ}$ defines a smooth 1-form on $Q$, and
\item[c)] $\gamma$, $\Omega$ and $H$ do not depend on the $V$-coordinate, i.e., $L_{\partial_V}\Omega =0$,  $L_{\partial_V}\gamma =0$, and $\partial_VH =0$ (so in particular, $\partial_V$ is a Killing - indeed parallel - null vector field).
\end{itemize}
Time orientation is chosen so that $\partial_V$ is past time-oriented. If $\Omega \equiv 0$, then $(\tilde{M},\tilde{g})$ is called a {\em generalized $pp$-wave}.
\end{definition}

Although this definition has been phrased in a somewhat pedantic way for the sake of precision, the reader will immediately recognize that it is simply meant to give a global version of the local structure that any Brinkmann space $(M,g)$ possesses (cf. Eq. (\ref{localBrink})). Minguzzi \cite{minguzzi1,minguzzi2} studied in detail a number of geometric properties of standard Brinkmann spacetimes, and in particular their causal properties. They have the important feature of being stably causal if and only if they are strongly causal (cf. Theorem 4.6 in Ref. \cite{minguzzi1}). This is a situation analogue to the case of stationary spacetimes, which are causally continuous if and only if they are distinguishing (cf. Proposition 3.1 of \cite{JS}). However, although a standard Brinkmann spacetime is always causal, it need not even be distinguishing (cf. Proposition 2.1 of \cite{FS}), unlike a standard stationary spacetime.

\begin{remark}
\label{key}
{\em With the notation as in Definition \ref{BEspacetimes}, a standard Brinkmann spacetime has the following features.
\begin{enumerate}
\item[1)] There exists a complete null parallel vector field, namely $X=\partial_V$;
\item[2)] There exists a complete vector field conjugate to $\partial_V$, namely $Y=\partial_U$;
\item[3)] There exists a complete {\em null} vector field conjugate to $\partial_V$, namely $\hat{Y} = \partial_U - H \partial_V$;
\item[4)] There exists a slice for the flow of $\partial_V$, namely any $V = const.$ surface. Note that $\partial_U$ is tangent to this surface.
\item[5)] As observed before, a standard Brinkmann is causal, so the action of $\partial_V$ is free. By $(4)$ and Proposition \ref{contractible}, it is also proper.
\item[6)] The joint $\mathbb{R}^2$-action that $\partial_V$ and $\partial_U$ generate is free and has any $V= const.,U =const.$ codimension 2 submanifold as a slice, so again by Proposition \ref{contractible}, it is also proper.
\end{enumerate}
}
\end{remark}

Next, we want to globalize the standard Brinkmann structure that every Brinkmann space locally has. Remark \ref{key} indicates that one cannot expect to be able to do this without completeness assumptions for a conjugate vector field, as well as for $X$. Indeed, a vector field conjugate to $X$ may or may not be complete. As a simple example, let $M= \{(u,v,y) \in \mathbb{R}^3 \,|\, -1 < v < 1 \}$ with metric given by $ds^2 = 2du(dv + dy) + dy^2$. If we take $X = \partial_u$ then $X$ is a complete null parallel vector field, while $Y = \partial_v$ and $Z=\partial_y -\frac{1}{2}\partial_u$ are both null vector fields conjugate to $X$. Note that $Y$ is incomplete, but $Z$ is complete.

This motivates the following definition.

\begin{definition}
\label{transversal}
A Brinkmann spacetime $(M,g)$ with complete parallel null vector field $X$ is {\em transversally complete} if there exists a complete vector field $Y \in \Gamma(TM)$ conjugate to $X$. We say that the $\mathbb{R}$-action $\rho$ generated by $X$ is {\em null-polar} if there exist a slice $\Sigma$ for $\rho$ together with a complete vector field $V \in \Gamma (T\Sigma)$ tangent to $\Sigma$ such that $g(X,V) =1$ everywhere on $\Sigma$.
\end{definition}
%\begin{definition}
%\label{transversal}
%Let $(M,g)$ be a Brinkmann spacetime with a complete parallel null vector field $X$.  The $\mathbb{R}$-action $\rho$ generated by $X$ is {\em transversally complete} if there exists a complete vector field $Y \in \Gamma(TM)$ conjugate to $X$. We will say that $\rho$ is {\em null-polar} if there exist a slice $\Sigma$ for $\rho$ together with a complete vector field $V \in \Gamma (T\Sigma)$ tangent to $\Sigma$ such that $g(X,V) =1$ everywhere on $\Sigma$.
%\end{definition}

\begin{remark}
{\em The term ``null-polar'' is motivated by the following null analogue of polar actions in Riemannian geometry. Assume that $(M,g)$ admits a null hypersurface $S \subset M$ with complete null geodesic generators which is a slice for the action $\rho$. To fix ideas, take the null parallel vector field $X$ to be past-directed.  $S$ is in particular closed. It is well-known that $S$ will have a smooth tangent future-directed null vector field $K$, unique up to multiplication by a smooth positive function, whose integral curves are pregeodesics whose images are precisely the null geodesic generators. Fix the normalization of $K$ such that $g(X,K) =1$. Then
\[
\nabla _K K = \lambda K
\]
for some $\lambda \in C^{\infty}(S)$. Now,
\[
\lambda = \lambda g(X,K) = g(X,\nabla _K K) = - g(\nabla_K X, K) =0,
\]
since $X$ is parallel. Hence, $K$ is a geodesic vector field, and thus complete. We conclude that $\rho$ is null-polar.}
\end{remark}

Actually, these notions are not independent for causal Brinkmann spacetimes, as the following result shows.
\begin{proposition}
\label{implication}
Let $(M,g)$ be a causal Brinkmann spacetime with a complete parallel null vector field $X$. If the flow $\rho$ of $X$ is null-polar, then $(M,g)$ is transversally complete.
\end{proposition}
{\em Proof.} Let $\Sigma \subset M$ a slice, together with a complete vector field $V \in \Gamma (T\Sigma)$ tangent to $\Sigma$ such that $g(X,V) =1$ everywhere on $\Sigma$. By Proposition \ref{contractible}, $\rho$ is proper and free, and $\rho|_{\Sigma}: \mathbb{R} \times \Sigma \rightarrow M$ is a diffeomorphism. Let $\tilde{Y}$ be the lift of $V$ to $\mathbb{R}\times \Sigma$. Then, the vector field $Y = (\rho|_{\Sigma})_{\ast} \tilde{Y}$ is complete, and arises from lifting $V$ along the orbits of $X$, so that $[X,Y]=0$. Due to this and since $X$ is Killing, $g(X,Y)=1$, so that $Y$ is a complete conjugate vector field as desired.
\qcd

Standard Brinkmann spacetimes has both features: it is transversally complete and the action of $\partial_V$ is null-polar. Indeeed, with mild causal assumptions these properties characterize these spacetimes.
\begin{theorem}
\label{basicsplitting}
Let $(M,g)$ be a Brinkmann spacetime with a complete parallel null vector field $X$ such that either
\begin{itemize}
\item[i)] $(M,g)$ is causal and the flow $\rho$ of $X$ is null-polar, or
\item[ii)] $(M,g)$ is strongly causal and transversally complete.
\end{itemize}
Then, the universal covering spacetime $(\overline{M}, \overline{g})$ of $(M,g)$ is isometric to a standard Brinkmann spacetime. The isometry can be chosen to be such that it associates the lift $\overline{X}$ of $X$ with $\partial_V$.
\end{theorem}
{\em Proof.} Note that since $(M,g)$ is causal (resp. strongly causal), so is its universal cover $(\overline{M}, \overline{g})$. Alternative (i) implies, via Proposition \ref{implication}, that $(M,g)$ is transversally complete, so we may assume this to be the case anyway. Moreover, either item (i) or (ii) ensure that the action $\rho$ is free and proper (cf. Proposition \ref{contractible} and Theorem \ref{jose}). The lift $\overline{X}$ of $X$ to $\overline{M}$ is still a complete null parallel vector field thereon. The action $\overline{\rho}$ generated by $\overline{X}$ on $\overline{M}$ is still free and proper (since $(\overline{M}, \overline{g})$ is causal and the lift of a slice for $\rho$ is clearly a slice for $\overline{\rho}$). In addition, the lift $\overline{Y}$ of a complete vector field $Y \in \Gamma(TM)$ conjugate to $X$ is also a complete vector field on $\overline{M}$ conjugate to $\overline{Y}$. Therefore, for the remainder of the proof, we will just drop the overline and assume that $M$ is itself simply connected. In that case, according to Lemma \ref{simple}, $X$ is gradient, and we pick $u \in C^{\infty}(M)$ for which $X = \nabla u$.

Fix the complete conjugate vector field $Y$, and denote its flow by $\phi$. Now, for a given $p \in M$, consider the integral curves $\alpha_p: t \in \mathbb{R} \mapsto \rho_t(p) \in M$ of $X$ and $\beta_p: s \in \mathbb{R} \mapsto \phi_s(p) \in M$ of $Y$ through $p$. Then we have
\[
(u \circ \alpha_p)'(t) = g_{\rho_t(p)}(\nabla u(\rho_t(p)),X_{\rho_t(p)}) = g_{\rho_t(p)}(X_{\rho_t(p)},X_{\rho_t(p)}) \equiv 0,
\]
i.e., $u$ is constant along the orbits of $\rho$. Similarly,
\[
(u \circ \beta_p)'(s) = g_{\phi_s(p)}(\nabla u(\phi_s(p)),Y_{\phi_s(p)}) \equiv 1,
\]
i.e.,
\begin{equation}
\label{floweffect}
u(\phi_s(p)) = u(p) + s,
\end{equation}
$\forall s \in \mathbb{R}$, and thus $u$ is surjective along the orbit of $p$ by $\phi$, i.e., $u(\mathbb{R}p) = \mathbb{R}$.

Let ${\cal N} := u^{-1}(0)$. Since $X=\nabla u$ is everywhere non-zero, ${\cal N}$ is a smooth embedded null hypersurface in $(M,g)$. Now, on the one hand, Proposition 8 of \cite{leistner} implies that the mapping
\[
\varphi: (s,p) \in \mathbb{R} \times {\cal N} \mapsto \phi_s(p) \in M
\]
is a diffeomorphism. On the other hand, ${\cal N}$ is invariant by the action $\rho$ of $X$. Therefore, given any slice (in the sense of Proposition \ref{contractible}, and which in the present case is a hypersurface) $\Sigma \subset M$, and for each $p \in \Sigma\cap {\cal N}$, $X(p) \in T_p{\cal N} \setminus T_p\Sigma$, that is, the intersection $\Sigma \cap {\cal N}$ is transversal, and hence the restricted action $\rho|_{\cal N}$ is proper by Corollary \ref{corcontractible}. We conclude that ${\cal N}$ is an $\mathbb{R}$ principal bundle over the quotient $Q:= {\cal N}$, and in particular ${\cal N} \simeq \mathbb{R} \times Q$. Fixing a global trivialization $\tau : \mathbb{R} \times Q \rightarrow {\cal N}$ of this bundle, define
\[
\Phi: (t,s,x) \in \mathbb{R}^2 \times Q \mapsto \phi_s(\tau(t,x)) \in M.
\]
Since $\Phi = \varphi \circ i \circ \tau$, where $i: {\cal N} \hookrightarrow M$ is the inclusion, this map is easily seen to be a diffeomorphism.

We adopt here the same notation as in Definition \ref{BEspacetimes}: $V,U: \mathbb{R}^2 \times Q\rightarrow \mathbb{R}$ denote the standard projection onto the first and second slots of $\mathbb{R}^2$, respectively. We note first that
\begin{equation}
\label{checkU}
u(\Phi(t,s,x)) = u(\phi_{s}(\tau(t,x))) = u(\tau(s,x)) + s = s = U(t,s,x)
\end{equation}
for all $t, s \in \mathbb{R}$ and all $x \in Q$, where we have used (\ref{floweffect}) for the second equality.
In other words, $u \circ \Phi = U$.

Second, we claim that
\begin{equation}
\label{checkV}
\Phi_{\ast}(\partial_V) = X \mbox{ and } \Phi_{\ast}(\partial_U) = Y.
\end{equation}
Indeed, given $(t_0,s_0,x_0) \in \mathbb{R}^2 \times Q$ and $f \in C^{\infty}(M)$, we compute:
\begin{eqnarray}
[(d \Phi)_{(t_0,s_0,x_0)}(\partial_V)](f) &=& (\partial_V(f \circ \Phi))(t_0,s_0,x_0) \\ \nonumber
           &=& \lim _{t \rightarrow 0} \frac{f(\phi_{s_0}(\tau(t_0 + t,x_0))) -f(\phi_{s_0}(\tau(t_0,x_0)))}{t} \\ \nonumber
           &=& \lim _{t \rightarrow 0} \frac{f(\phi_{s_0}(\rho _t(\tau(t_0,x_0)))) -f(\phi_{s_0}(\tau(t_0,x_0)))}{t}\\ \nonumber
           &=&  \lim _{t \rightarrow 0} \frac{f(\rho_{t}(\phi _{s_0}(\tau(t_0,x_0)))) -f(\phi_{s_0}(\tau(t_0,x_0)))}{t}\\ \nonumber
 &=&  \lim _{t \rightarrow 0} \frac{f(\rho_{t}(\Phi(t_0,s_0,x_0))) -f(\Phi(t_0,s_0,x_0))}{t}\\ \nonumber
           &=& [X(f)](\Phi(t_0,s_0,x_0)),
\end{eqnarray}
where we have used the fact that $\tau$ is a principal bundle trivialization on the third equality, and the fact that the actions of $X$ and $Y$ must commute (since $[X,Y] =0$) on the fourth one. A similar (and even easier) calculation shows the other part of the claim.

Now, by Proposition \ref{conjugateexistence}, $\hat{Y}:= Y - \frac{1}{2} g(Y,Y)X$ is a null vector field also conjugate to $X$, so we can apply Proposition \ref{nulllemma} to get
\begin{equation}
\label{firstmetric}
g = \hat{\omega} \otimes du + du \otimes \hat{\omega} + \hat{g},
\end{equation}
where $\hat{\omega} = g(\hat{Y}, \,. \,)$ is the conjugate 1-form associated with $\hat{Y}$. Thus, we can {\em define} the metric on $\mathbb{R}^2 \times Q$ as $\tilde{g}:= \Phi ^{\ast} g$, so that $\Psi$ becomes an isometry. Then, Eq. (\ref{firstmetric}) becomes
\begin{equation}
\label{secondmetric}
\tilde{g} = (\Phi ^{\ast} \hat{\omega}) \otimes dU + dU \otimes (\Phi ^{\ast} \hat{\omega}) + \gamma,
\end{equation}
where we have used Eq. (\ref{checkU}) and defined $\gamma := \Phi ^{\ast} \hat{g}$. Note that Eqs. (\ref{checkV}) imply
\begin{equation}
\label{checkomega}
L_{\partial_V} (\Phi ^{\ast} \hat{\omega}) = \Phi ^{\ast} (L_X \hat{\omega}) =0,
\end{equation}
since $\hat{\omega}$ is conjugate. Similarly, since $X$ is Killing, $L_{\partial_V} \tilde{g} =0$, whence we conclude that $L_{\partial_V} \tilde{g} =0$.

We can write
\[
\Phi ^{\ast} \hat{\omega} = \alpha dV + H dU + \Omega,
\]
for suitable $\alpha, H \in C^{\infty}(\mathbb{R}^2 \times Q)$ and $\Omega \in \Omega ^1 (\mathbb{R}^2 \times Q)$ that do not depend on the coordinate $V$ due to (\ref{checkomega}). In fact
\[
\alpha = \Phi ^{\ast} \hat{\omega}(\partial_V) = \hat{\omega}(\Phi_{\ast} \partial_V) = \hat{\omega}(X) =1,
\]
due to (\ref{checkV}). Again,
\[
H = \Phi ^{\ast} \hat{\omega}(\partial_U) = \hat{\omega}(\Phi_{\ast}(\partial_U)) = \hat{\omega}(Y)\circ \Phi = \frac{1}{2}g(Y,Y) \circ \Phi,
\]
where we have again used (\ref{checkV}). Hence, we finally get
\[
\tilde{g} = (dV + H dU + \Omega) \otimes dU + dU \otimes (dV + H dU + \Omega) + \gamma .
\]
\qcd

\begin{remark}
{\em Alternative (i) in Theorem \ref{basicsplitting} gives a precise null analogue of part (i) of the Javaloyes-S\'{a}nchez splitting theorem according to Proposition \ref{JavaloyesMiguel}. But to the best of our knowledge, there is NO analogue of part (ii) therein. In other words, {\em there is no causal condition on $(M,g)$ that might by itself ensure that the action of $X$ is null-polar}. Indeed, the existence of a complete conjugate vector field is not guaranteed even in globally hyperbolic spacetimes. To see this, consider the following example. In $3-d$ Minkowski spacetime $(\mathbb{R}^3,2dudv + dy^2)$, consider $M = \{ (v,u,y) \in \mathbb{R}^3 \, | \, u <1 \}$ with $g$ the restricted metric and time-orientation so that $\partial_v$ is past directed, and take $X= \partial_v$ restricted to $M$ as the pertinent complete null parallel vector field. It is easy to check that since $M$ is a past set in a globally hyperbolic spacetime, $(M,g)$ is itself globally hyperbolic. Any vector field conjugate to $X=\partial_v$ must be of the form
\[
Y = \partial_u + a(u,y) \partial_v + b(u,y) \partial_y.
\]
Now, pick the integral curve $\alpha(t) = (v(t),u(t),y(t))$ through $(0,0,0)$. Then $u(t) = t<1$, and hence $Y$ must be incomplete.}
\end{remark}

Our next and final result in this section gives a concrete situation in which Theorem \ref{basicsplitting} applies.

\begin{corollary}
Let $(M,g)$ be a strongly causal Brinkmann spacetime with a complete parallel null vector field $X$. Suppose that the orbit space of the associated $\mathbb{R}$-action $\rho$ is compact. (This will occur, for example, if $(M,g)$ is globally hyperbolic with a compact Cauchy hypersurface.) Then the universal covering spacetime $(\overline{M}, \overline{g})$ of $(M,g)$ is isometric to a standard Brinkmann spacetime. The isometry can be chosen to be such that it associates the lift $\overline{X}$ of $X$ with $\partial_V$.
\end{corollary}
{\em Proof.} Since $(M,g)$ is strongly causal, $\rho$ is free and proper (Theorem \ref{jose}). By the proof of Proposition \ref{contractible}, we can pick a slice $\Sigma$ for $\rho$ which is compact since it is diffeomorphic to the orbit space. Given any vector field $V \in \Gamma(T\Sigma)$ normalized so that $g(X,V) =1$, the compactness of $\Sigma$ will ensure it is complete, so that $\rho$ is null-polar. The result now follows from Theorem \ref{basicsplitting} (i).
\qcd

\section{A Rigidity conjecture for gravitational plane waves}\label{0.6}

As we have commented in the Introduction, a pp-wave is a standard Brinkmann spacetime $(M,g)$ such that $\Omega \equiv 0$ and $h_{ij} \equiv \delta_{ij}$ globally \cite{BE,EK}. This family of spacetimes has been the focus of intensive studies, both in the mathematical and physical literatures, since they provide idealized models for gravitational waves in General Relativity. The special case where H is quadratic and harmonic correspond to the important subclass of {\em gravitational plane waves}, which might be considered as idealized limiting configurations at ``infinite distance'' from gravitational radiating sources. The role played by these spaces in this context seems analogous to the one played by Minkowski space when it is considered as the asymptotic configuration associated to an isolated gravitational source.

Now, note that both ``models at infinity'' are geodesically complete. Geodesic incompleteness of causal geodesics have been used for a long time as a criterion for gravitational collapse in Physics \cite{BE,HE,oneill}. On physical grounds one would expect that geodesic complete solutions to, say, the Einstein field equation should be rare in a suitable sense. However, mathematically rigorous statements behind these considerations are far from obvious. A important illustration in the stationary case is formulated below, and was given in \cite{anderson}:
\begin{theorem}\label{a} (\cite[Theorem 0.1]{anderson}) Every geodesically complete, chronological, Ricci-flat $4$-dimensional stationary spacetime is isometric to (a quotient of) flat Minkowski spacetime.
\end{theorem}
In view of this theorem, it becomes natural to conceive of an analogous result for gravitational plane waves. Note that an interesting way of viewing Brinkmann spacetimes is as null analogues of stationary spacetimes. So, the null version of previous Anderson's rigidity theorem could be formulated as follows:
\begin{conjecture}\label{c} Every strongly causal, Ricci-flat $4$-dimensional Brinkamnn spacetime satisfying certain completeness condition is isometric to (a quotient of) a gravitational plane wave spacetime.
\end{conjecture}
Here, ``certain completeness condition'' play the role of geodesic completenes in Anderson's result. It is not clear to the authors if geodesic completeness is also sufficient in this case, but we believe that geodesic and transversal completeness (recall Definition \ref{transversal}) ought to be enough. On the other hand, strong causality here replaces chronology in Anderson's result, since, for the null case, a little more causality is required to ensure that the quotient associated to the Killing vector field behaves well. Since every plane wave spacetime is causallty continuous (thus strongly causal), this is not very restrictive.

Theorem \ref{basicsplitting} becomes a first, but very important, step in the proof of Conjecture \ref{c}, since it allows one to restrict one's considerations to standard Brinkmann spacetimes. In fact, together with J. Herrera, we have been able to show \cite[Theorem 3.1]{CFH} that, under the hypotheses of the conjecture, a standard Brinkmann space must be a pp-wave. So, the proof of Conjecture \ref{c} actually reduces to show that {\em any geodesically complete strongly causal Ricci-flat $4$-dimensional pp-wave is a plane wave}. This statement essentially coincides with the problem formulated by Ehlers-Kundt in \cite[Section 2-5.7]{EK}, which has been cited or studied by different authors in the last years \cite{B}, \cite{HR}, \cite{FlS}, \cite{leistner}, \cite{CFH}.

 A key result \cite{CFH} is precisely the partial solution to the Ehlers-Kundt and to the Conjecture \ref{c} given as follows.
 \begin{theorem} (\cite[Theorem 2.1]{CFH}) Let $(M,g)$ be a geodesically complete strongly causal Ricci-flat $4$-dimensional Brinkmann spacetime with a complete parallel vector field $X$. Assume also that there exists a Killing vector field conjugate to $X$. Then, $(M,g)$ is (a quotient of) a plane wave.
\end{theorem}

\section{Acknowledgements}
The authors are partially supported by the Spanish MICINN-FEDER Grant MTM2013-47828-C2-2-P, with FEDER funds.
IPCS wishes to acknowledge a research scholarship from CNPq, Brazil, Programa Ci�ncias sem Fronteira, process number 200428/2015-2. He is very grateful to the staff and faculty members of the Department of Mathematics of the University of the Miami for the kind hospitality while this work was being finished. JLF is also supported by the Regional J. Andaluc\'{i}a Grant PP09-FQM-4496, with FEDER funds. He gratefully acknowledges the hospitality of the Department of Mathematics of Universidade Federal de Santa Catarina, where part of this work was developed.

%%%%%%%%%%%%%%%%%%%%%%%%%%%%%%%%%%%%%%%%%%%%%%%%%%%%%%%%%%%%%%%%%%%%%%%%%%%%%%%%%%%%%
%References
%%%%%%%%%%%%%%%%%%%%%%%%%%%%%%%%%%%%%%%%%%%%%%%%%%%%%%%%%%%%%%%%%%%%%%%%%%%%%%%%%%%%%

\vspace*{\fill}

%-------------------------------------------------------------------------------------------------------------

\vspace{.5cm}


\begin{thebibliography}{1}
%
\bibitem{anderson}
M.T. Anderson, {\em On stationary vacuum solutions to the Einstein equations}, Ann. H. Poincar\'{e} {\bf 1} (2000), 977-994.
%
\bibitem{BCFl} R. Bartolo, A.M. Candela, J.L. Flores, {\em Connections by geodesics on globally hyperbolic spacetimes with a lightlike Killing vector field}, Rev. Mat. Ibero., in press.
%
\bibitem{batat}
W. Batat, G. Calvaruso and B. De Leo, {\em Homogeneous Lorentzian 3-manifolds
with a parallel null vector field}, Balkan J. Geom. Appl. {\bf 14} (2009) 11-20.
%
\bibitem{baum}
H. Baum, K. L\"{a}rz and T. Leistner, {\em On the full holonomy group of special Lorentzian manifolds}, Math. Z. {\bf 277} (2014) 797-828.
%
\bibitem{BE}
J.K. Beem, P.E. Ehrlich and K.L. Easley
  {\it Global Lorentzian Geometry}, $2^{\rm{nd}}$ ed.,
  Marcel Dekker, New York (1996).
%
\bibitem{BS}
A.N. Bernal and M. S\'{a}nchez, {\em Smoothness of time function and the metric splitting of globally hyperbolic spacetimes}, Commun. Math. Phys. {\bf 257} (2005) 43-50.
%
\bibitem{B} J. Bicak: Selected solutions of Einstein's field equations: Their role in general relativity and astrophysics. In: {\em Einstein's Field Equations and Their Physical Interpretations, Lect. Notes Phys. {\bf 540}, ed by Schmidt (Springer, Heidelberg 2000) pp 1-126. gr-qc/0004016.}
%
\bibitem{brinkmann}
H.W. Brinkmann, {\em Einstein spaces which are mapped conformally on each other}, Math. Ann. {\bf 94} (1925) 119-145.
%
\bibitem{Josetal}
A.M. Candela, J.L. Flores, M. S\'{a}nchez, {\em Global hyperbolicity and Palais�Smale condition for action functionals in stationary spacetimes}, Advances in Mathematics, {\bf 218} (2008) 515-536.
%
\bibitem{caponio}
E. Caponio, M.A. Javaloyes and A. Masiello, {\em On the energy functional on Finsler manifolds and applications to stationary spacetimes}, Math. Ann., {\bf 351} (2011) 365-392.
%
\bibitem{CFH} I. Costa e Silva, J.L. Flores, J. Herrera, {\em Rigidity of geodesic completeness in the Brinkmann class of gravitational wave spacetimes}. Preprint (2016).
%
%\bibitem{duval1}
%C. Duval, G. Burdet, H. P. K\'{a}nzle and M. Perrin, {\em Bargmann structures and Newton-Cartan
%theory} Phys. Rev. D {\bf 31} (1985) 1841-1853.
%
%\bibitem{duval2}
%C. Duval, G. Gibbons and P. Horv\'{a}thy, {\em Celestial mechanics, conformal structures, and
%gravitational waves}, Phys. Rev. D {\bf 43} (1991) 3907-3922.
%
\bibitem{duggal}
K. Duggal and B. Sahin, {\em Differential Geometry of Lightlike Submanifolds}, Birkh�user, Basel (2010).
%
\bibitem{EK}
J. Ehlers and K. Kundt, {\em Exact solutions of the gravitational field equations}, in {\em Gravitation: an introduction to current research}, L. Witten (ed.), J. Wiley \& Sons, New York (1962) pp. 49-101.
%
\bibitem{galaev}
A.S. Galaev and T. Leistner, {\em On the local structure of Lorentzian Einstein manifolds with parallel distribution of null lines}, Class. Quantum Grav. {\bf 27} (2010) 225003.
%
%\bibitem{eisenhart}
%L.P. Eisenhart, {\em Dynamical trajectories and geodesics}, Ann. Math. {\bf 30} (1929) 591-606.
%
\bibitem{FS}
J.L. Flores and M. S\'{a}nchez, {\em Causality and
conjugate points in general plane waves}, Class Quantum Grav. {\bf 20} (2003)
2275-2291.
%
\bibitem{FlS}
J.L. Flores and M. S\'{a}nchez, On the geometry of pp-wave type spacetimes. In {\em Analytical and
numerical approaches to mathematical relativity}, Lecture Notes in Phys. {\bf 692}, p.
79�98. Springer, Berlin, 2006.
%
\bibitem{geroch1}
R.P. Geroch, {\em A method for generating solution of Einstein's equations}, J. Math. Phys. {\bf 12} (1971) 918-924.
%
\bibitem{geroch2}
R.P. Geroch, {\em A method for generating solution of Einstein's equations 2}, J. Math. Phys. {\bf 13} (1972) 394-404.
%
\bibitem{Harris}
S.G. Harris, {\em Conformally stationary spacetimes}, Class. Quantum Grav. {\bf 9} (1992) 1823-1827.
%
\bibitem{HL}
S.G. Harris and R.J. Low, {\em Causal monotonicity, omniscient foliations and the shape of space}, Class. Quantum Grav. {\bf 18} (2001) 27-43.
%
\bibitem{HE}
S.W. Hawking and G.F.R. Ellis,
  {\it The Large Scale Structure of Space-time},
  Cambridge University Press, Cambridge (1973).
%
\bibitem{heusler}
M. Heusler, {\em Black Hole Uniqueness Theorems},  Cambridge University Press, Cambridge (1996).
%
\bibitem{HR} V.E. Hubeny, M. Rangamani, {\em Causal Structures of pp-waves}, J. High Energy Phys. {\bf 0212}, 043 (2002).
%
\bibitem{JS}
M.A. Javaloyes and M. S\'{a}nchez, {\em A note on the existence of standard splittings for conformally stationary spacetimes}, Class. Quantum Grav. {\bf 25} (2008) 168001.
%
\bibitem{KN}
S. Kobayashi and K. Nomizu, {\em Foundations of Differential Geometry}, Vol. 1, Wiley, New York (1962).
%
\bibitem{Lee}
J.M. Lee, {\em Introduction to smooth manifolds}, Springer, New York (2003).
%
\bibitem{leistner}
T. Leistner and D. Schliebner, {\em Completeness of compact Lorentzian manifolds with abelian holonomy}, Math. Ann., DOI 10.1007/s00208-015-1270-4 (2015).
%
\bibitem{minguzzi1}
E. Minguzzi, {\em Causality of spacetimes admitting a parallel null vector and weak KAM theory}, ArXiv: 1211.2685 (2012).
%
\bibitem{minguzzi2}
E. Minguzzi, {\em Eisenhart's theorem and the causal simplicity of Eisenhart's
spacetime} Class. Quantum Grav. {\bf 24} (2007) 2781-2807.
%
\bibitem{MS}
E. Minguzzi and M. S\'{a}nchez, {\em The causal hierarchy of spacetimes}, in {\em Recent Developments in Semi-Riemannian Geometry}, ESI Lect. Math. Phys., eds. H. Baum, D. Alekseevsky, Eur. Math. Soc. Publ. House, Zurich (2008).
%
\bibitem{oneill}
B. O'Neill,
  {\it Semi-Riemannian Geometry with Applications to Relativity},
  Academic Press, New York (1983).
%
\bibitem{Palais}
R. Palais, {\em On the existence of slices for actions of non-compact Lie groups}, Ann. Math. {\bf 73} (1961) 295-323.
%
\bibitem{Penrose}
R. Penrose, {\em Singularities and time-asymmetry}, in {\em General Relativity: An Einstein Centenary Survey}, eds. S.W. Hawking and W. Israel, Cambridge University Press, Cambridge (1979).
%
\bibitem{penrose}
R. Penrose, {\em A remarkable property of plane waves in general relativity},
Rev. Mod. Phys. {\bf 37} (1965) 215-220.
%
\bibitem{piccionezeghib}
P. Piccione and A. Zeghib, {\em Actions of discrete groups on stationary Lorentz manifolds}, Ergod. Th. Dynam. Sys. {\bf 34} (2014) 1640-1673.
%
\bibitem{Vandenban}
E.P. van den Ban, {\em Lecture notes on Lie groups}, available at http://www.staff.science.uu.nl/~ban00101/lecnotes/lie2010.pdf.
%
\bibitem{wald} R.M. Wald, {\em General Relativity}, Universidty of Chicago Press, Chicago, IL, 1984.
%
\bibitem{walker}
A.G. Walker, {\em Canonical form for a Riemannian space with a parallel field of null planes}, Quarterly J. Math. Oxford {\bf 1} (1950) 69-79.
%
\bibitem{zeghib}
A. Zeghib, {\em Isometry groups and geodesic foliations of Lorentz manifolds. I. Foundations of Lorentz dynamics}, Geom. Func. Anal. {\bf 9} (1999) 823-854.



\end{thebibliography}
\end{document}